%% file: main.tex
\pdfoutput=1
\documentclass[10pt, logo, twocolumn, copyright]{nv}

\usepackage{graphicx}
\definecolor{nvidiagreen}{HTML}{76B900}

\input{preamble}

\usepackage{mdframed}
\usepackage{color}
\usepackage{xcolor}
\usepackage[utf8]{inputenc} 
\usepackage[T1]{fontenc}    

\usepackage{amsfonts}       
\usepackage{nicefrac}       
\usepackage{microtype}      
\usepackage{multirow}
\usepackage{multicol}
\usepackage{tabto}
\usepackage{xspace}
\usepackage{amsmath}
\usepackage{adjustbox}
\usepackage{enumitem}
\usepackage{wrapfig}
\usepackage{dblfloatfix}
\usepackage{times}
\usepackage{verbatim}
\usepackage{amssymb}
\usepackage{mathtools}
\usepackage{caption}
\usepackage{subcaption}
\usepackage{array}
\usepackage{colortbl}
\usepackage{booktabs}
\usepackage{bbm}
\usepackage{makecell}
\usepackage{float}
\usepackage{siunitx}
\usepackage{pifont}
\usepackage{marvosym}
\usepackage{listings}
\usepackage{pdflscape}
\usepackage{footmisc}
\usepackage{url}
\usepackage{tabularx}
\usepackage{arydshln}
\usepackage{hhline}
\usepackage{diagbox}
\usepackage{tcolorbox}
\usepackage[nameinlink]{cleveref}
\usepackage{hyperref}
\usepackage[square,sort,comma,numbers]{natbib} 
\usepackage{fp} 
\usepackage{authblk}
\usepackage{xspace}

\crefname{section}{Sec.}{Sec.}
\crefname{proposition}{Proposition.}{Proposition.}
\crefname{equation}{Eq.}{Eqs.}
\crefname{figure}{Fig.}{Figs.}
\crefname{table}{Tab.}{Tabs.}
\crefname{algorithm}{Algorithm}{Algorithms}
\crefname{appendix}{Appendix}{Appendices}
\Crefname{thm}{Thm}{Thm}
\definecolor{codebg}{RGB}{245, 245, 245} 
\definecolor{keywordcolor}{RGB}{0, 0, 153} 
\definecolor{commentcolor}{RGB}{34, 139, 34} 
\definecolor{stringcolor}{RGB}{163, 21, 21}
\definecolor{numbercolor}{RGB}{128, 128, 128}

\def\eg{\emph{e.g.}}

\title{
\model: One-Step Diffusion with Continuous-Time Consistency Distillation
}

\author{
\centering
\fontsize{10pt}{18pt}\selectfont
Junsong Chen\textsuperscript{1$*$} ~~
Shuchen Xue\textsuperscript{5$*$} ~~
Yuyang Zhao\textsuperscript{1$\dagger$} ~~
Jincheng Yu\textsuperscript{1$\dagger$} ~~
Sayak Paul\textsuperscript{4} ~~
\\
\vspace{0em}
\textbf{\fontsize{10pt}{18pt}\selectfont
Junyu Chen\textsuperscript{3} ~~ 
Dongyun Zou\textsuperscript{3} ~~
Han Cai\textsuperscript{1} ~~
Song Han\textsuperscript{1,2} ~~
Enze Xie\textsuperscript{1} ~~ 
}
\\
\vspace{2mm}
{\normalsize \textsuperscript{1}NVIDIA ~~
\textsuperscript{2}MIT ~~
\textsuperscript{3}Tsinghua University ~~
\textsuperscript{4}Huggingface ~~
\textsuperscript{5}Independent Researcher} \\
\vspace{2pt}
{\footnotesize $^*$Equal contribution. $^\dagger$Core contribution.}
}

\begin{abstract}\small
\textbf{Abstract:} This paper presents \model{}, an efficient diffusion model for ultra-fast text-to-image (T2I) generation. \model is built on a pre-trained foundation model and augmented with hybrid distillation, dramatically reducing inference steps from 20 to 1-4.
We introduce three key innovations:
\textbf{(1)} We propose a training-free approach that transforms a pre-trained flow-matching model for continuous-time consistency distillation (sCM), eliminating costly training from scratch and achieving high training efficiency. Our hybrid distillation strategy combines sCM with latent adversarial distillation (LADD): sCM ensures alignment with the teacher model, while LADD enhances single-step generation fidelity.
\textbf{(2)} \model is a unified step-adaptive model that achieves high-quality generation in 1-4 steps, eliminating step-specific training and improving efficiency.
\textbf{(3)} We integrate ControlNet with \model for real-time interactive image generation, enabling instant visual feedback for user interaction.
\model{} establishes a new Pareto frontier in speed-quality tradeoffs, achieving state-of-the-art performance with 7.59 FID and 0.74 GenEval in only 1 step — outperforming FLUX-schnell (7.94 FID / 0.71 GenEval) while being 10× faster (0.1s vs 1.1s on H100). It also achieves 0.1s~(T2I) and 0.25s~(ControlNet) latency for 1024$\times$1024 images on H100, and 0.31s~(T2I) on an RTX 4090, showcasing its exceptional efficiency and potential for AI-powered consumer applications (AIPC).
Code and pre-trained models will be open-sourced.
    \newline
    \textbf{Links:} \hspace{2pt}
    {\hypersetup{urlcolor=nvidiagreen}
    \href{https://github.com/NVlabs/Sana}{GitHub Code} |
    \href{https://huggingface.co/collections/Efficient-Large-Model/sana-sprint-67d6810d65235085b3b17c76}{HF Models} |
    \href{https://nvlabs.github.io/Sana/Sprint/} {Project Page}
    }
\end{abstract}

\begin{document}
\maketitle

\input{sec/1_intro}
\input{sec/2_preliminaries}
\input{sec/3_methods}
\input{sec/4_exps}
\input{sec/6_related_works}
\input{sec/5_conclusion}

\input{sec/8_acknowledgment}

\newpage
\appendix
\onecolumn

\input{sec/7_appendix}

\clearpage

{
  \small
  \bibliographystyle{unsrt}
  \bibliography{main}
}

\end{document}

%% file: preamble.tex
\usepackage{amsmath}
\usepackage{amssymb}
\usepackage{mathtools}
\usepackage{amsthm}
\usepackage{multirow}
\usepackage{algorithm}
\usepackage{algpseudocode}
\usepackage{url}            %
\usepackage{booktabs}       %
\usepackage{amsfonts}       %
\usepackage{nicefrac}       %
\usepackage{microtype}      %
\usepackage{xcolor}         %
\usepackage{mathtools}
\usepackage{listings}

\usepackage{xcolor}
\definecolor{pearDark}{RGB}{34,139,34}  
\definecolor{mygreen}{RGB}{34,139,34}
\definecolor{mylightblue}{RGB}{0,162,230}
\definecolor{deepyellow}{RGB}{255,215,0}
\definecolor{nvgreen}{RGB}{118, 185, 0}

\usepackage{xspace}
\newcommand{\model}{SANA-Sprint\xspace}

\newcommand{\bx}{\boldsymbol{x}}
\newcommand{\bz}{\boldsymbol{z}}
\newtheorem{proposition}{Proposition}[section]

\newtheorem*{remark}{Remark}


%% file: sec/1_intro.tex
\section{Introduction}
\label{sec:intro}

\input{Figures_tex/teaser}

The computational intensity of diffusion generative models~\citep{ho2020denoising,song2020score}, typically requiring 50-100 iterative denoising steps, has driven significant innovation by time-step distillation to enable efficient inference. Current methodologies primarily coalesce into two dominant paradigms: (1) distribution-based distillations like GAN~\citep{goodfellow2014generative} (e.g., ADD~\citep{sauer2024adversarial}, LADD~\citep{sauer2024fast}) and its variational score distillation (VSD) variants~\citep{poole2022dreamfusion,wang2023prolificdreamer,luo2023diff} leverage joint training to align single-step outputs with multi-step teacher's distributions, and (2) trajectory-based distillations like Direct Distillation~\citep{luhman2021knowledge}, Progressive Distillation~\citep{salimans2022progressive,meng2023distillation}, Consistency Models (CMs)~\citep{song2023consistency} (e.g. LCM~\citep{luo2023latent}, CTM~\citep{kim2023consistency}, MCM~\citep{heek2024multistep}, PCM~\citep{wang2024phased}, sCM~\citep{lu2024simplifying}) learn ODE solution across reduced sampling intervals. Together, these methods achieve 10-100× image generation speedup while maintaining competitive generation quality, positioning distillation as a critical pathway toward practical deployment.

Despite their promise, key limitations hinder broader adoption. GAN-based methods suffer from training instability due to oscillatory adversarial dynamics and mode collapse. GANs face challenges due to the need to map noise to natural images without supervision, making unpaired learning more ill-posed than paired learning, as highlighted in~\citep{zhu2017unpaired,kang2024distilling}. This instability is compounded by architectural rigidity, which demands meticulous hyperparameter tuning when adapting to new backbones or settings. VSD-based methods involve the joint training of an additional diffusion model, which increases computational overhead and imposes significant pressure on GPU memory, and requires careful tuning~\citep{yin2024improved}. Consistency models, while stable, suffer quality erosion in ultra-few-step regimes (e.g., $<$4 steps), particularly in text-to-image tasks where trajectory truncation errors degrade semantic alignment. These challenges underscore the need for a distillation framework that harmonizes efficiency, flexibility, and quality.

In this work, we present SANA-Sprint, an efficient diffusion model for one-step high-quality text-to-image (T2I) generation.
Our approach builds on a pre-trained image generation model SANA and recent advancements in continuous-time consistency models (sCMs)~\citep{lu2024simplifying}, preserving the benefits of previous consistency-based models while mitigating the discretization errors of their discrete-time counterparts. 
To achieve the one-step generation, we first transform SANA, a Flow Matching model, to the TrigFlow model, which is required for sCM distillation, through a lossless mathematical transformation. 
Then, to mitigate the instability of distillation, we adapt the QK norm in self- and cross-attention in SANA along with dense time embeddings to allow efficient knowledge transfer from the pre-trained models without retraining the teacher model. We further combine sCM with LADD's adversarial distillation to enable fast convergence and high-fidelity generation while retaining the advantages of sCMs. 
Note that, although validated primarily on SANA, our method can benefit other mainstream flow-matching models such as FLUX~\citep{FLUX} and SD3~\citep{esser2024scaling}. 

As a result, \model achieves excellent speed/quality tradeoff, benefiting from a hybrid objective, inheriting sCM's diversity preservation and alignment with the teacher, while integrating LADD's fidelity enhancement: experiments show a 0.6 lower FID and 0.4 higher CLIP-Score at 2-step generations compared to standalone sCM, with 3.9 lower FID and 0.9 higher CLIP-Score over pure latent adversarial approaches.
As shown in~\cref{fig:teaser}, SANA-Sprint achieves state-of-the-art performance in FID and GenEval benchmark, surpassing recent advanced methods including SD3.5-Turbo, SDXL-DMD2, and Flux-schnell. Especially, \model is 64.7$\times$ faster than Flux-Schnell and exceeds in FID~(7.59 vs 7.94) and GenEval~(0.74 vs 0.71).

Moreover, SANA-Sprint demonstrates unprecedented inference speed—generating 1024$\times$1024 images in 0.31 seconds on a laptop with consumer-grade GPUs (NVIDIA RTX 4090) and 0.1 seconds on H100 GPU, 8.4$\times$ speedup than teacher model SANA. This efficiency unlocks transformative applications that require instant visual feedback: in ControlNet-guided image generation/editing, by integrating with ControlNet, \model enables instant interaction with 250ms latency on H100. \model exhibits robust scalability and is potentially suitable for human-in-the-loop creative workflows, AIPC, and immersive AR/VR interfaces.

\noindent In summary, our key contributions are threefold: 

\begin{itemize}  
    \item \textbf{Hybrid Distillation Framework}:
    We designed an innovative hybrid distillation framework that seamlessly transforms the flow model into the Trigflow model, integrating continuous-time consistency models (sCM) with latent adversarial diffusion distillation (LADD). This framework leverages sCM's diversity preservation and alignment with the teacher alongside LADD's fidelity enhancement, enabling unified step-adaptive sampling.

    \item \textbf{Excellent Speed/Quality Tradeoff}: \model achieves exceptional performance with only 1-4 steps. \model generates a 1024×1024 image in only 0.10s-0.18s on H100, achieving state-of-the-art 7.59 FID on MJHQ-30K and 0.74 GenEval score - surpassing FLUX-schnell (7.94 FID/0.71 GenEval)  while being 10$\times$ faster.
    
    \item \textbf{Real-Time Interactive Generation}: By integrating ControlNet with \model, we enable real-time interactive image generation in 0.25s on H100. This facilitates immediate visual feedback in human-in-the-loop creative workflows, enabling better human-computer interaction. 
\end{itemize}

%% file: Figures_tex/teaser.tex
\begin{figure*}[th]
    \centering
    \includegraphics[width=0.97\linewidth]{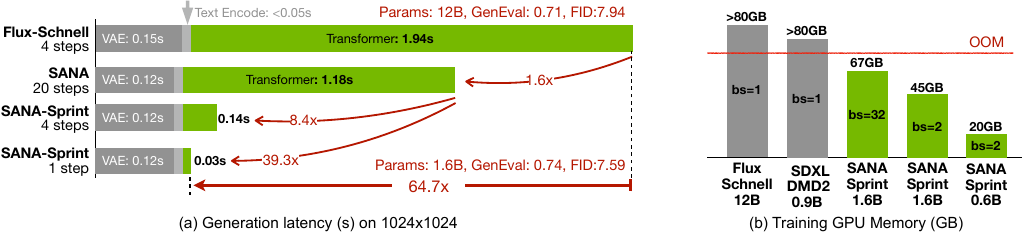}
    \vspace{-0.5em}
    \caption{(a) Our \model accelerate the inference speed for generating 1024 $\times$ 1024 images, achieving a remarkable speedup from FULX-Schnell's 1.94 seconds to only 0.03 seconds. This represents a 64$\times$ improvement over the current state-of-the-art step-distilled model, FLUX-Schnell, as measured with a batch size of 1 on an NVIDIA A100 GPU. The ratio is calculated based on Transformer latency.
    (b) Additionally, our model demonstrates efficient GPU memory usage during training, outperforming other distillation methods in terms of memory cost. The GPU memory is measured using official code, 1024 $\times$ 1024 images and on a single A100 GPU.
    }
    \label{fig:teaser}
    \vspace{-1em}
\end{figure*}

%% file: sec/2_preliminaries.tex
\section{Preliminaries}

\subsection{Diffusion Model and Its Variants}

Diffusion models~\citep{ho2020denoising,song2020score} diffuse clean data sample $\bx_0\sim p_{data}$ from data distribution to noisy data $\bx_t = \alpha_t \bx_0 + \sigma_t \bz$, where $t\in [0,T]$ represents time within the interval, $\bz \sim \mathcal{N}(\mathbf{0}, \boldsymbol{I})$ is a standard Gaussian noise. The terminal distribution $p_T$ of $\bx_T$ exactly or approximately follows a Gaussian distribution. Typically, diffusion models train a noise prediction network $\boldsymbol{\epsilon_\theta}$ using $\mathbb{E}_{\bx_0, \bz, t}[\| \boldsymbol{\epsilon_\theta}(\bx_t, t) - \bz \|^2]$, which is equivalent to denoising score matching loss~\citep{vincent2011connection,song2020score}. The sampling process of diffusion models involves solving the probability flow ODE (PF-ODE)~\citep{song2020score} $\frac{\mathrm{d} \bx_t}{\mathrm{d} t} = \frac{\mathrm{d} \log \alpha_t}{\mathrm{d} t} \bx_t + (\frac{\mathrm{d}  \sigma_t}{\mathrm{d} t} - \frac{\mathrm{d} \log \alpha_t}{\mathrm{d} t} \sigma_t) \boldsymbol{\epsilon_\theta}(\bx_t, t)$ with the initial value $\bx_1 \sim \mathcal{N}(\mathbf{0}, \boldsymbol{I})$. Below, we will introduce two recent formulations of diffusion models that have received significant attention.

Flow Matching~\citep{peluchetti2022nondenoising,liu2022flow,lipman2022flow} considers a linear interpolation noising process by defining $\alpha_t = 1-t, \sigma_t = t, T = 1$. The flow matching models train a velocity prediction network $\boldsymbol{v_\theta}$ using $\mathbb{E}_{\bx_0, \bz, t}[w(t)\| \boldsymbol{v_\theta}(\bx_t, t) - (\bz-\bx_0) \|^2]$, where $w(t)$ is a weighting function. The sampling of flow models solves the PF-ODE $\frac{\mathrm{d} \bx_t}{\mathrm{d} t} = \boldsymbol{v_\theta}(\bx_t, t)$ with the initial value $\bx_1 \sim \mathcal{N}(\mathbf{0}, \boldsymbol{I})$.

TrigFlow~\citep{albergo2022building,lu2024simplifying} considers a spherical linear interpolation noising process by defining $\alpha_t = \cos(t), \sigma_t = \sin(t), T = \frac{\pi}{2}$. Moreover, Trigflow assumes the noise $\bz \sim \mathcal{N}(\mathbf{0}, \sigma_d^2\boldsymbol{I})$, where $\sigma_d$ represents the standard deviation of data distribution $p_{data}$. The TrigFlow models train a velocity prediction network $\boldsymbol{F_{\theta}}$ using $\mathbb{E}_{\bx_0, \bz, t}[w(t)\| \sigma_d \boldsymbol{F_{\theta}}(\frac{\bx_t}{\sigma_d}, t) - (\cos(t)\bz-\sin(t)\bx_0) \|^2]$, where $w(t)$ is a weighting function. The sampling of TrigFlow models solves the PF-ODE defined by $\frac{\mathrm{d} \bx_t}{\mathrm{d} t} = \sigma_d \boldsymbol{F_{\theta}}(\frac{\bx_t}{\sigma_d}, t)$ starting from $\bx_{\frac{\pi}{2}} \sim \mathcal{N}(\mathbf{0}, \sigma_d^2\boldsymbol{I})$.

Diffusion, Flow Matching, and TrigFlow are all continuous-time generative models that differ in their interpolation schemes and velocity field parameterizations.

\input{Figures_tex/paradigm}

\subsection{Consistency Models}

A consistency model (CM)~\citep{song2023consistency} parameterizes a neural network $\boldsymbol{f_{\theta}}(\bx_t, t)$ which is trained to predict the solution $\bx_0$ of the PF-ODE, which is the terminal clean data along the trajectory of the PF-ODE (regardless of its position in the trajectory), starting from the noisy observation $\bx_t$. The conventional approach parameterizes the CM using skip connections, bearing a close resemblance to~\citep{karras2022elucidating,balaji2022ediff}
\begin{equation}\label{eq:skip}
\boldsymbol{f_{\theta}}(\bx_t, t) = c_{\text{skip}}(t)\bx_t + c_{\text{out}}(t)\boldsymbol{F_{\theta}}(\bx_t, t),
\end{equation}
where $c_{\text{skip}}(t)$ and $c_{\text{out}}(t)$ are differentiable functions satisfying $c_{\text{skip}}(0) = 1$ and $c_{\text{out}}(0)=0$ to ensure the boundary conditions $\boldsymbol{f_{\theta}}(\bx_0, 0) = \bx_0$. $\boldsymbol{F_{\theta}}$ indicates the pretrained diffusion/flow model and $\boldsymbol{f_{\theta}}$ is the data prediction model.
Based on the training approach, CMs can be categorized into two types: 
discrete-time~\citep{song2023consistency,luo2023latent} and continuous-time~\citep{song2023consistency,lu2024simplifying}.

\noindent Discrete-time CMs are trained with the following objective
\begin{equation}\label{eq:disc cm loss}
l_{CM}^{\Delta t} = \mathbb{E}_{\bx_t, t}[d(\boldsymbol{f_{\theta}}(\bx_t, t), \boldsymbol{f_{\theta^-}}(\bx_{t-\Delta t}, t-\Delta t))],
\end{equation}
where $\boldsymbol{\theta^-}$ is the $\texttt{stopgrad}$ version of $\boldsymbol{\theta}$, $w(t)$ is the weighting function, $\Delta t$ is a small time interval, and $\bx_{t-\Delta t}$ is obtained from $\bx_t$ by running a numerical ODE solver. $d(\cdot,\cdot)$ is a metric such as $\ell_1$, squared $\ell_2$, Pseudo-Huber loss, and the LPIPS loss~\citep{zhang2018unreasonable}.

\noindent Although discrete-time CMs work well in practice, the additional discretization errors brought by numerical ODE solvers are inevitable. Continuous-time CMs correspond to the limiting case of $\Delta t \rightarrow 0$ in \cref{eq:disc cm loss}. When choosing $d(\bx,\boldsymbol{y})=\|\bx-\boldsymbol{y}\|_2^2$, the expression simplifies to:
\begin{equation}\small
l_{CM}^{cont.} := \lim_{\Delta t \rightarrow 0}\frac{l_{CM}^{\Delta t}}{\Delta t} =  \mathbb{E}_{\bx_t, t}\left[w(t) \langle \boldsymbol{f_{\theta}}(\bx_t, t),  \frac{\mathrm{d} \boldsymbol{f_{\theta^-}}}{\mathrm{d} t}(\bx_t, t) \rangle\right],
\label{eq:continuous cm loss}
\end{equation}
where $\frac{\mathrm{d} \boldsymbol{f_{\theta^-}}(\bx_t, t)}{\mathrm{d} t} = \frac{\partial \boldsymbol{f_{\theta^-}}(\bx_t, t)}{\partial t} + \nabla_{\bx_t} \boldsymbol{f_{\theta^-}}(\bx_t, t) \frac{\mathrm{d}\bx_t}{\mathrm{d}t}$. The infinitesimal step of $\frac{\mathrm{d}\bx_t}{\mathrm{d}t}$ replaces numerical ODE solvers, thereby eliminating discretization errors.

Specifically, under TrigFlow where $c_{\text{skip}}(t) = \cos(t)$ and $c_{\text{out}}(t) = -\sin(t)$, sCM's parameterization and arithmetic coefficients are simplified to the following form:

\begin{equation}
\label{eq:para scm}
\boldsymbol{f_{\theta}}(\bx_t, t) = \cos(t) \bx_t - \sin(t) \sigma_d \boldsymbol{F_{\theta}}(\frac{\bx_t}{\sigma_d}, t),
\end{equation}
and the time derivative becomes:
\begin{equation}
\label{eq:jvp}
\begin{aligned}
\frac{\mathrm{d} \boldsymbol{f_{\theta^-}}(\bx_t, t)}{\mathrm{d} t} =& -\cos(t)\left( \sigma_d \boldsymbol{F_{\theta^-}}(\frac{\bx_t}{\sigma_d}, t) -  \frac{\mathrm{d}\bx_t}{\mathrm{d}t} \right)\\
&-\sin(t)\left( \bx_t + \sigma_d \frac{\mathrm{d} \boldsymbol{F_{\theta^-}}(\frac{\bx_t}{\sigma_d}, t)}{\mathrm{d} t} \right).\\
\end{aligned}
\end{equation}

%% file: Figures_tex/paradigm.tex
\begin{figure*}[th]
    \centering
    \includegraphics[width=0.88\linewidth]{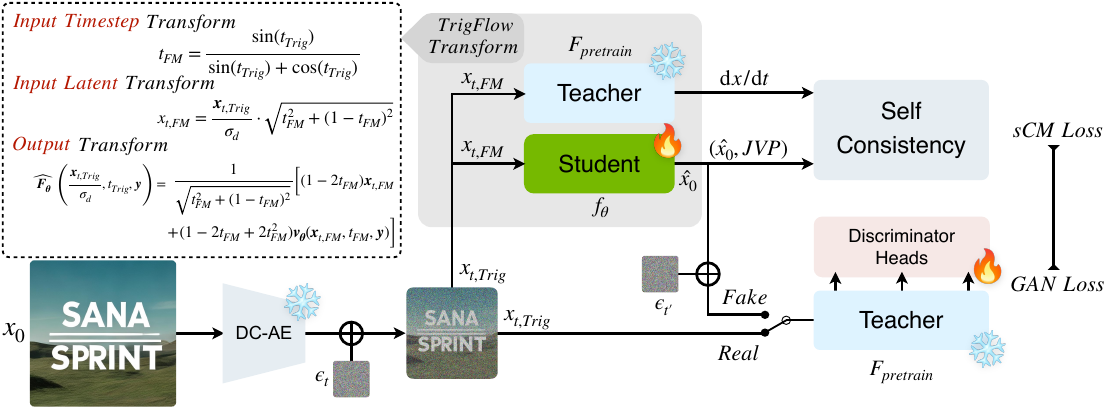}
    \vspace{-0.5em}
    \caption{
    \textbf{Training paradigm of \model{}.} In \model{}, we use the student model for synthetic data generation~($\hat{x_0}$) and $\text{JVP}$ calculation, and we use the teacher model for velocity~($\mathrm{d}x/\mathrm{d}t$) compute and its feature for the GAN loss, which allows us train sCM and GAN together and have only one training model purely in the latent space. Details of training objective and \textit{TrigFlow Transformation} are in \cref{eq:scm loss}, \cref{eq:gan generator loss} and Sec.~\ref{Sec:trans}.
    }
    \label{fig:paradigm}
    \vspace{-1em}
\end{figure*}

%% file: sec/3_methods.tex
\section{Method}

sCM~\citep{lu2024simplifying} simplify continuous-time CMs using the TrigFlow formulation. While this provides an elegant framework, most score-based generative models are based on diffusion or flow matching formulations. One possible approach is to develop separate training algorithms for continuous-time CMs under these formulations, but this requires distinct algorithm designs and hyperparameter tuning, increasing complexity. Alternatively, one could pretrain a dedicated TrigFlow model, as in \citep{lu2024simplifying}, but this significantly increases computational cost.

To address these challenges, we propose a simple method to transform a pre-trained flow matching model into a TrigFlow model through straightforward mathematical input and output transformations. This approach makes it possible to strictly follow the training algorithm in \citep{lu2024simplifying}, eliminating the need for separate algorithm designs while fully leveraging existing pre-trained models. The transformation process for general diffusion models can be carried out in a similar manner, which we omit here for simplicity.

\subsection{Training-Free Transformation to TrigFlow}
\label{Sec:trans}
Score-based generative models (diffusion, flow matching, and TrigFlow) can denoise data with proper data scales and signal-to-noise ratios (SNRs)\footnote{For a diffusion model $\bx_t = \alpha_t\bx_0 + \sigma_t \boldsymbol{z}$, SNR is defined as $\frac{\alpha_t^2}{\sigma_t^2}$ } aligned with training. However, flow matching cannot directly denoise TrigFlow-scheduled data due to three mismatches: First, their time domains differ: TrigFlow uses $[0, \frac{\pi}{2}]$, while flow matching is defined on $[0,1]$. Second, their noise schedules are distinct—TrigFlow maintains $\cos^2(t_{\texttt{Trig}}) + \sin^2(t_{\texttt{Trig}}) = 1$, while flow matching yields $t_{\texttt{FM}}^2 + (1 - t_{\texttt{FM}})^2 < 1$, causing data scale discrepancies. Finally, their prediction targets differ: flow matching predicts $z - x_0$ with static coefficients $(1,-1)$, whereas TrigFlow predicts $\cos(t)z - \sin(t)x_0$ with time-varying coefficients. These mismatches in temporal parameterization, SNR, and output necessitate explicit input/output transformations.

To clarify, we use the subscript $_{\texttt{Trig}}$ to denote noisy data under the TrigFlow framework and $_{\texttt{FM}}$ to denote noisy data under the flow matching framework. The following proposition outlines the transformation from flow matching models to TrigFlow models, which is theoretically lossless. 

\begin{remark}
We prioritize seamlessly transforming existing noise schedules, \eg flow matching, into TrigFlow while integrating the sCM framework with minimal modifications. This approach avoids the need for pre-training a dedicated TrigFlow model, as in \cite{lu2024simplifying}, although it involves a deviation from the unit variance principle in~\citep{karras2022elucidating,lu2024simplifying}.
\end{remark}

\begin{proposition}\label{prop:trans}
 Given a noisy data $\frac{\bx_{t,\texttt{Trig}}}{\sigma_d}$ under TrigFlow noise schedule, a flow matching model can denoise it via $\boldsymbol{v_{\theta}}(\bx_{t, \texttt{FM}}, t_{\texttt{FM}}, \boldsymbol{y})$, where
\begin{equation}\small\label{eq:t trans}
t_{\texttt{FM}} = \frac{\sin{(t_{\texttt{Trig}})}}{\sin{(t_{\texttt{Trig}})} + \cos{(t_{\texttt{Trig}})}},
\end{equation}
\begin{equation}\small\label{eq:x trans}
x_{t,\texttt{FM}} = \frac{\bx_{t,\texttt{Trig}}}{\sigma_d}\cdot\sqrt{t_{\texttt{FM}}^2 + (1-t_{\texttt{FM}})^2} .
\end{equation}

Given $\boldsymbol{v_{\theta}}(\bx_{t, \texttt{FM}}, t_{\texttt{FM}}, \boldsymbol{y})$, the best estimator for the TrigFlow model $\boldsymbol{F_{\theta}}$ is the following:
\begin{equation}
\small
\label{eq:output trans}
\begin{aligned}
&\widehat{\boldsymbol{F_{\theta}}}\left(\frac{\bx_{t,\texttt{Trig}}}{\sigma_d}, t_{\texttt{Trig}}, \boldsymbol{y}\right) \\
= &\frac{1}{\sqrt{t_{\texttt{FM}}^2 + (1-t_{\texttt{FM}})^2}} \Big[
  (1-2t_{\texttt{FM}}) \bx_{t, \texttt{FM}} \\
  &\quad + (1-2t_{\texttt{FM}} + 2t_{\texttt{FM}}^2) \boldsymbol{v_{\theta}}(\bx_{t, \texttt{FM}}, t_{\texttt{FM}}, \boldsymbol{y})
\Big].
\end{aligned}
\end{equation}
Furthermore, the transformation is lossless in theory. 
\end{proposition}

The details and proof of \cref{prop:trans} are in~\cref{app:proof}. The transformations of both input and output are all differentiable making it compatible with auto differentiation. As validated by \cref{tab:lossless-conversion}, the transformation is lossless in both theory and practice. The training-free transformation is depicted in the gray box of~\cref{fig:paradigm}.

\input{Figures_tex/timestep_qknorm_grad}

\input{Tables/trigflow_flow_teacher_metrics}

\paragraph{Self Consistency Loss.}

With the lossless transformation established, we can seamlessly adopt the training algorithm and pipeline of sCM without other modification. This allows us to directly follow the sCM training framework. Our final sCM loss is the following:
\begin{equation}\label{eq:scm loss}
\small
\begin{aligned}
&\mathcal{L}_{\text{sCM}}(\boldsymbol{\theta}, \boldsymbol{\phi}) = \mathbb{E}_{\bx_t, t}\Big[\frac{e^{w_{\boldsymbol{\phi}}(t)}}{D}\Big\| 
\widehat{\boldsymbol{F_{\theta}}}\left(\frac{\bx_{t}}{\sigma_d}, t, \boldsymbol{y}\right) \\
&- \widehat{\boldsymbol{F_{\theta^-}}}\left(\frac{\bx_{t}}{\sigma_d}, t, \boldsymbol{y}\right)- \cos(t) \frac{\mathrm{d} \widehat{\boldsymbol{f_{\theta^-}}}(\bx_t, t, y)}{\mathrm{d} t} \Big\|_2^2 - w_{\boldsymbol{\phi}}(t) \Big]  \\
\end{aligned}
\end{equation}
where $\widehat{\boldsymbol{f_{\theta^-}}}$ is the parameterized sCM as in \cref{eq:para scm} after replacing $\boldsymbol{F_{\theta}}$ with $\widehat{\boldsymbol{F_{\theta}}}$ in \cref{prop:trans}, $\bx_t, t$ refers to $\bx_{t,\texttt{Trig}}, t_{\texttt{Trig}}$, and $w_{\boldsymbol{\phi}}(t)$ is an adaptive weighting function to minimize variance across different timesteps following~\citep{karras2024analyzing,lu2024simplifying}.

\subsection{Stabilizing Continuous-Time Distillation}
\label{stabilizing continuous-time DiTs}

To stabilize continuous-time consistency distillation, we address two key challenges: training instabilities and excessively large gradient norms that occur when scaling up the model size and increasing resolution, leading to model collapse. We achieve this by refining the time-embedding to be denser and integrating QK-Normalization into self- and cross-attention mechanisms. These modifications enable efficient training and improve stability, allowing for robust performance at higher resolutions and larger model sizes.
\\
\\
\noindent \textbf{Dense Time-Embedding.}
As analyzed in sCM~\citep{lu2024simplifying}, the instability issues in continuous-time CMs primarily stem from the unstable scale of $\frac{\mathrm{d}\boldsymbol{f_{\theta}}}{\mathrm{d}t}$ in~\cref{eq:scm loss}. This instability can be traced back to the expression $\frac{\mathrm{d}F_{\theta^-}}{\mathrm{d}t} = \nabla_{\bx_t} F_{\theta^-} \frac{\mathrm{d}\bx_t}{\mathrm{d}t} + \partial_t F_{\theta^-}$ in~\cref{eq:jvp}, which ultimately originates from the time derivative term $\partial_t F_{\theta^-}$:
\begin{equation}\small
\partial_t F_{\theta^-} = \frac{\partial F_{\theta^-}}{\partial \text{emb}(c_{\text{noise}})} \cdot \frac{\partial \text{emb}(c_{\text{noise}})}{\partial c_{\text{noise}}} \cdot \boldsymbol{\frac{\partial c_{\text{noise}}(t)}{\partial t}}
\end{equation}

In previous flow matching models like SANA~\citep{xie2024sana}, SD3~\citep{esser2024scaling}, and FLUX~\citep{FLUX}, the noise coefficient $c_{\text{noise}}(t) = 1000t$ amplifies the time derivative $\partial_t F_{\theta^-}$ by a factor of 1000, leading to significant training fluctuations. To address this, we set $c_{\text{noise}}(t) = t$ and fine-tuned SANA for 5k iterations. As shown in~\cref{fig:timestep_qknorm_grad}~(b), this adjustment reduce excessively large gradient norms (originally exceeding $10^3$) to more stable levels. Furthermore, PCA visualization in~\cref{fig:timestep_qknorm_grad}~(c) reveals that our dense time-embedding design results in more densely packed and similar embeddings for timesteps between 0$\sim$1. This refinement improved training stability and accelerated convergence over 15k iterations.

\noindent \textbf{QK-Normalization.} When scaling up the model from 0.6B to 1.6B, we encounter similar issues with excessively large gradient norms, often exceeding $10^3$, which lead to training collapse. To address this, we introduce RMS normalization~\citep{zhang2019root} to the Query and Key in both self- and cross-attention modules of the teacher model during fine-tuning. This modification enhances training stability significantly, even with a brief fine-tuning process of only 5,000 iterations. By using the fine-tuned teacher model to initialize the student model, we achieve a more stable gradient norm, as shown in~\cref{fig:timestep_qknorm_grad}~(a), thereby making the distillation process viable where it was previously infeasible.

\input{Figures_tex/compare-other-methods}

\subsection{Improving Continuous-Time CMs with GAN}\label{sec:ladd}

CTM~\citep{kim2023consistency} analyzes that CMs distill teacher information in a local manner, where at each iteration, the student model learns from local time intervals. 
This leads the model to learn cross timestep information under the implicit extrapolation, which can slow the convergence speed.
To address this limitation, we introduce an additional adversarial loss~\citep{sauer2024fast} to provide direct global supervision across different timesteps, improving both the convergence speed and the output quality.

GANs~\citep{goodfellow2014generative} consist of a generator $G$ and a discriminator $D$ that compete in a zero-sum game to produce realistic synthetic data. Diffusion-GANs~\citep{wang2022diffusion} and LADD~\citep{sauer2024fast} extend this framework by enabling the discriminator to distinguish between noisy real and fake samples. Furthermore, LADD introduces a novel approach by utilizing a frozen teacher model as a feature extractor and training multiple discriminator heads on the teacher model. This methodology facilitates direct adversarial supervision in the latent space, as opposed to the traditional pixel space, leading to more efficient and effective training. Following LADD, we use a hinge loss~\citep{lim2017geometric} to train the student model and discriminator
\begin{equation}\small\label{eq:gan generator loss}
\begin{aligned}
&\mathcal{L}_{\text{adv}}^G(\boldsymbol{\theta})   \\
&=-\mathbb{E}_{\bx_0, s, t}\Big[  \sum_{k}  D_{\boldsymbol{\psi}, k}(\boldsymbol{F}_{\boldsymbol{\theta}^{\texttt{pre}}, k}(\hat{\bx}_{s}^{\boldsymbol{f_{\theta}}}, s, \boldsymbol{y})) \Big],
\end{aligned}
\vspace{-1em}
\end{equation}
\begin{equation}\small
\begin{aligned}
&\mathcal{L}_{\text{adv}}^D(\boldsymbol{\psi})   \\
&=\mathbb{E}_{\bx_0, s}\Big[  \sum_{k} \text{ReLU}\Big(1- D_{\boldsymbol{\psi}, k}(\boldsymbol{F}_{\boldsymbol{\theta}^{\texttt{pre}}, k}(\bx_s, s, \boldsymbol{y}))\Big) \Big]  \\
&+ \mathbb{E}_{\bx_0, s, t}\Big[  \sum_{k} \text{ReLU}\Big(1+ D_{\boldsymbol{\psi}, k}(\boldsymbol{F}_{\boldsymbol{\theta}^{\texttt{pre}}, k}(\hat{\bx}_{s}^{\boldsymbol{f_{\theta^-}}}, s, \boldsymbol{y}))\Big) \Big],
\end{aligned}
\vspace{-1em}
\end{equation}\\
where $\bx_s, \hat{\bx}_{s}^{\boldsymbol{f_{\theta}}}, \hat{\bx}_{s}^{\boldsymbol{f_{\theta^-}}}$ are the noisy versions of $\bx_0, \hat{\bx}_{0}^{\boldsymbol{f_{\theta}}} := \boldsymbol{f_{\theta}}(\bx_t,t,\boldsymbol{y}), \hat{\bx}_{0}^{\boldsymbol{f_{\theta^-}}} := \boldsymbol{f_{\theta^-}}(\bx_t,t,\boldsymbol{y})$. 

The adversarial loss $\mathcal{L}_{\text{adv}}$ is equivalent to the GAN loss shown at the bottom of~\cref{fig:paradigm}. In summary, \model combines sCM loss with GAN loss: $\mathcal{L} = \mathcal{L}_{sCM} + \lambda \mathcal{L}_{\text{adv}}$, where $\lambda = 0.5$ by default, as in~\cref{tab:loss_weight}.

\paragraph{Additional Max-Time Weighting.} In our early experiments, we adopt the timestep sampling distribution of sCM’s generator (student model) for GAN loss, given by  $t=\arctan{(\frac{e^\tau}{\sigma_d})}$, where $\tau\sim\mathcal{N}(P_{\text{mean}}, P_{\text{std}}^2)$ with two hyperparameters $P_{\text{mean}}$ and $P_{\text{std}}$. To further enhance one- and few-step generation performance and improve overall generation quality, we introduce an additional weighting at $t = \frac{\pi}{2}$. Specifically, with probability $p$, the training timestep is set to $\frac{\pi}{2}$, while with probability $1-p$, it follows the original timestep sampling distribution of sCM’s generator. We find that this modification significantly improves the model’s capability for one- and few-step generation, as shown in~\cref{tab:maxT_ladd}.

\input{Tables/main_results}

\subsection{Application: Real-Time Interactive Generation} 
Extending the \model to image-to-image tasks is straightforward. We apply the \model training pipeline to ControlNet~\cite{zhang2023adding} tasks, which utilize both images and prompts as instructions. Our approach involves continuing the training of a pre-trained text-to-image diffusion model with a diffusion objective on a dataset adjusted for ControlNet tasks, resulting in the SANA-ControlNet model. We then distill this model using the \model framework to obtain \model-ControlNet.

For ControlNet tasks, we extract Holistically-Nested Edge Detection (HED) scribbles from input images as conditions to guide image generation. Following PixArt's~\cite{chen2024pixartdelta} design principles, we train our SANA-ControlNet teacher model on 1024$\times$1024 resolution images. During sampling, HED maps serve as additional conditioning inputs to the Transformer model, allowing precise control over image generation while maintaining structural details.
Our experiments show that the distilled \model-ControlNet model retains the controllability of the teacher model and achieves fast inference speeds of approximately 200 ms on H100 machines, enabling near-real-time interaction. The effectiveness of our approach is demonstrated in~\cref{app:qual results}. 

%% file: Figures_tex/timestep_qknorm_grad.tex
\begin{figure*}[t]
    \centering
    \includegraphics[width=0.95\linewidth]{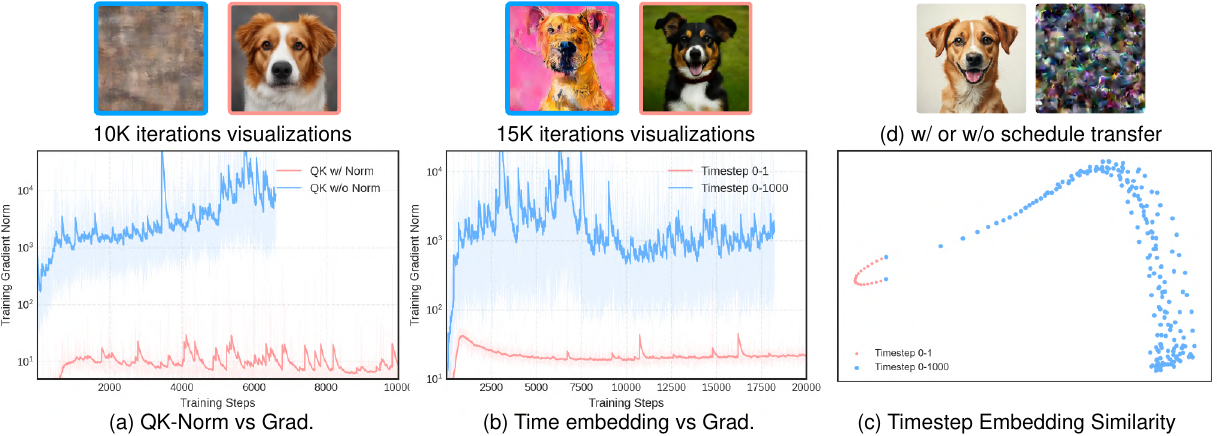}
    \vspace{-0.5em}
    \caption{
    \textbf{Efficient Distillation via QK Normalization, Dense Timestep Embedding, and Training-free Schedule Transformation.}
    (a) We compare gradient norms and visualizations with/without QK Normalization, showing its stabilizing effect.
    (b) Gradient norm curves for timestep scales (0$\sim$1 vs. 0$\sim$1000) highlight impacts on stability and stability and quality.
    (c) PCA-based similarity analysis of timestep embeddings.
    (d) Image results after 5,000 iterations of fine-tuning with (left) and without (right) the proposed schedule transfer (\cref{Sec:trans}).
    }
    \label{fig:timestep_qknorm_grad}
    \vspace{-1em}
\end{figure*}

%% file: Tables/trigflow_flow_teacher_metrics.tex
\begin{table}[ht]
    \centering
    \caption{\textbf{Comparison of original Flow-based SANA model and training-free transformation of TrigFlow-based \model{} model.} We evaluate the FID and CLIP-Score before and after the transformation in~\cref{Sec:trans}.}
    \vspace{-0.5em}
    \scalebox{0.9}{
    \begin{tabular}{l|c|c}
    \toprule
        \textbf{Method} & FID~$\downarrow$ & CLIP-Score~$\uparrow$ \\
    \midrule
        Flow Euler 50 steps & 5.81 & 28.810 \\
        TrigFlow Euler 50 steps & 5.73 & 28.806 \\
    \bottomrule
    \end{tabular}
    }
    \vspace{-1.6em}
    \label{tab:lossless-conversion}
\end{table}

%% file: Figures_tex/compare-other-methods.tex
\begin{figure*}[t]
    \centering
    \includegraphics[width=0.9\linewidth]{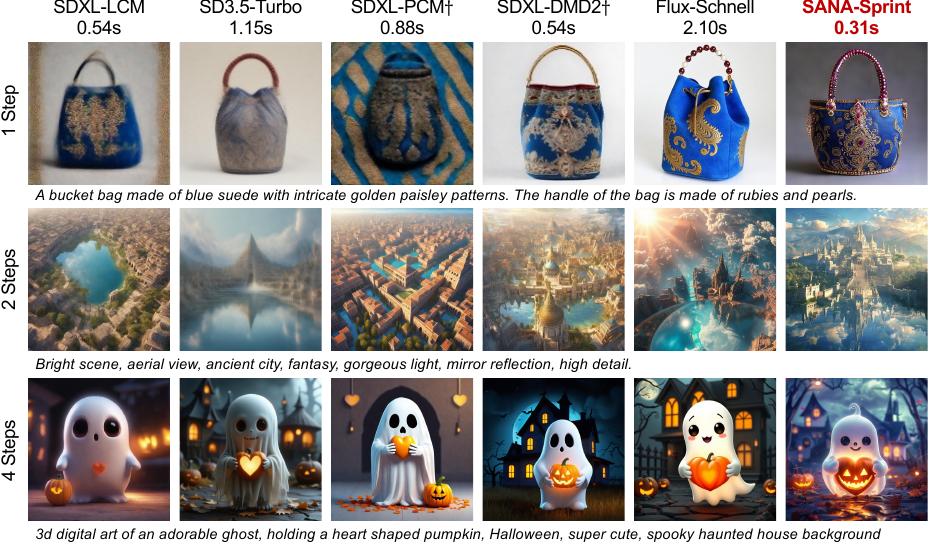}
    \vspace{-0.5em}
    \caption{\textbf{Visual comparison among \model and selected competing methods in different inference steps.} \dag~indicates that distinct models are required for different inference steps, and time below the method name is the latency of 4 steps tested on A100 GPU. \model produces images with superior realism and text alignment in all inference steps with the fastest speed.
    }
    \vspace{-1em}
    \label{fig:compare-other-methods}
\end{figure*}

%% file: Tables/main_results.tex
\begin{table*}[t]
\centering
\caption{\textbf{Comprehensive comparison of \model with SOTA approaches in efficiency and performance.} The speed is tested on one A100 GPU with BF16 Precision. Throughput: Measured with batch=10. Latency: Measured with batch=1. We highlight the \textbf{best} and \underline{second best} entries. \dag~indicates that distinct models are required for different inference steps.
}
\small
\vspace{-0.5em}
\label{tab:main_comparison}{
\scalebox{0.95}{
\begin{tabular}{l|l|ccc|cccc}
\toprule
\multicolumn{2}{c|}{\multirow{2}{*}{\textbf{Methods}}} & \textbf{Inference} & \textbf{Throughput} & \textbf{Latency} & \textbf{Params} & \multirow{2}{*}{\textbf{FID~$\downarrow$}} & \multirow{2}{*}{\textbf{CLIP~$\uparrow$}} & \multirow{2}{*}{\textbf{GenEval~$\uparrow$}} \\
\multicolumn{2}{c|}{} & \textbf{steps} & \textbf{(samples/s)} & \textbf{(s)} & \textbf{(B)} &  &  &  \\

\midrule

\multicolumn{1}{c|}{\multirow{7}{*}{\rotatebox{90}{\textbf{Pre-train Models}}}} & SDXL~\citep{podell2023sdxl}             & 50    & 0.15  & 6.5   & 2.6   & 6.63              & 29.03             & 0.55   \\
&PixArt-$\Sigma$~\citep{chenpixartsigma} & 20    & 0.4   & 2.7   & 0.6   & 6.15              & 28.26             & 0.54    \\
&SD3-medium~\citep{sd3}                  & 28    & 0.28  & 4.4   & 2.0   & 11.92             & 27.83             & 0.62   \\
&FLUX-dev~\citep{FLUX}                   & 50    & 0.04  & 23.0  & 12.0  & 10.15             & 27.47             & 0.67     \\
&Playground v3~\citep{liu2024playground} & -     & 0.06  & 15.0  & 24    & -                 & -                 & 0.76  \\
&SANA 0.6B~\citep{xie2024sana}           & 20    & 1.7   & 0.9   & 0.6   & 5.81              & 28.36             & 0.64   \\
&SANA 1.6B~\citep{xie2024sana}           & 20    & 1.0   & 1.2   & 1.6   & 5.76  & 28.67             & 0.66   \\

\midrule

\multicolumn{1}{c|}{\multirow{28}{*}{\rotatebox{90}{\textbf{Distillation Models}}}} & SDXL-LCM~\cite{luo2023latent}           & 4     & 2.27  & 0.54  & 0.9   & 10.81     & 28.10     & 0.53   \\
&PixArt-LCM~\cite{chen2024pixartdelta}   & 4     & 2.61  & 0.50  & 0.6   & 8.63      & 27.40     & 0.44   \\
&PCM~\cite{wang2024phased}\dag           & 4     & 1.95  & 0.88  & 0.9   & 15.55     & 27.53     & 0.56   \\
&SD3.5-Turbo~\cite{esser2024scaling}     & 4     & 0.94  & 1.15  & 8.0   & 11.97     & 27.35     & 0.72   \\
&SDXL-DMD2~\cite{yin2024improved}\dag    & 4     & 2.27  & 0.54  & 0.9   & \textit{6.82}      & \textbf{28.84}     & 0.60   \\
&FLUX-schnell~\citep{FLUX}               & 4     & 0.5   & 2.10  & 12.0  & 7.94      & \textit{28.14}     & \textit{0.71}  \\
\cmidrule{2-9}
&\textbf{\model{} 0.6B}                  & 4     & 5.34  & 0.32  & 0.6   & \textbf{6.48}      & \underline{28.45}     & \underline{0.76}  \\
&\textbf{\model{} 1.6B}                  & 4     & 5.20  & 0.31  & 1.6   & \underline{6.54}      & \underline{28.45}     & \textbf{0.77} \\
\cmidrule{2-9}
&SDXL-LCM~\cite{luo2023latent}           & 2     & 2.89  & 0.40  & 0.9   & 18.11     & 27.51     & 0.44   \\
&PixArt-LCM~\cite{chen2024pixartdelta}   & 2     & 3.52  & 0.31  & 0.6   & 10.33     & 27.24     & 0.42   \\
&SD3.5-Turbo~\cite{esser2024scaling}     & 2     & 1.61  & 0.68  & 8.0   & 51.47     & 25.59     & 0.53   \\
&PCM~\cite{wang2024phased}\dag           & 2     & 2.62  & 0.56  & 0.9   & 14.70     & 27.66     & 0.55   \\
&SDXL-DMD2~\cite{yin2024improved}\dag    & 2     & 2.89  & 0.40  & 0.9   & \textit{7.61}      & \textbf{28.87}     & 0.58   \\
&FLUX-schnell~\citep{FLUX}               & 2     & 0.92  & 1.15  & 12.0  & 7.75      & 28.25     & \textit{0.71}   \\
\cmidrule{2-9}
&\textbf{\model{} 0.6B}                  & 2     & 6.46  & 0.25  & 0.6   & \underline{6.54}      & \textit{28.40}      & \underline{0.76}  \\
&\textbf{\model{} 1.6B}                  & 2     & 5.68  & 0.24  & 1.6   & \textbf{6.50}      & \underline{28.45}     & \textbf{0.77}  \\
\cmidrule{2-9}
&SDXL-LCM~\cite{luo2023latent}           & 1     & 3.36  & 0.32  & 0.9   & 50.51     & 24.45     & 0.28   \\
&PixArt-LCM~\cite{chen2024pixartdelta}   & 1     & 4.26  & 0.25  & 0.6   & 73.35     & 23.99     & 0.41   \\
&PixArt-DMD~\cite{chenpixartsigma}\dag   & 1     & 4.26  & 0.25  & 0.6   & 9.59      & 26.98     & 0.45   \\
&SD3.5-Turbo~\cite{esser2024scaling}     & 1     & 2.48  & 0.45  & 8.0   & 52.40     & 25.40     & 0.51   \\
&PCM~\cite{wang2024phased}\dag           & 1     & 3.16  & 0.40  & 0.9   & 30.11     & 26.47     & 0.42   \\
&SDXL-DMD2~\cite{yin2024improved}\dag    & 1     & 3.36  & 0.32  & 0.9   & \underline{7.10}      & \textbf{28.93}     & 0.59   \\
&FLUX-schnell~\citep{FLUX}               & 1     & 1.58  & 0.68 & 12.0   & \textit{7.26}      & \underline{28.49}     & \textit{0.69}   \\
\cmidrule{2-9}
&\textbf{\model{} 0.6B}                  & 1     & 7.22  & 0.21  & 0.6   & \textbf{7.04}      & 28.04     & \underline{0.72}  \\
&\textbf{\model{} 1.6B}                  & 1     & 6.71  & 0.21  & 1.6   & 7.69      & \textit{28.27}     & \textbf{0.76} \\
\bottomrule
\end{tabular}}
\vspace{-1em}
}
\end{table*}

%% file: sec/4_exps.tex
\section{Experiments}

\subsection{Experimental Setup}

Our experiments employ a two-phase training strategy, with detailed settings and evaluation protocols outlined in~\cref{app:exp setup}. The teacher models are pruned and fine-tuned from the larger SANA-1.5 4.8B model~\cite{xie2025sana}, followed by distillation using our proposed training paradigm. We evaluate performance using metrics including FID, CLIP Score on the MJHQ-30K~\citep{li2024playground}, and GenEval~\cite{ghosh2023geneval}.

\subsection{Efficiency and Performance Comparison}
We compare \model with state-of-the-art text-to-image diffusion and timestep distillation methods in~\cref{tab:main_comparison} and~\cref{fig:compare-other-methods}. Our SANA-Sprint models focus on timestep distillation, achieving high-quality generation with 1-4 inference steps, competing with the 20-step teacher model, as shown in~\cref{fig:compare-step}. More details about the timestep setting are given in~\cref{appendix:more ablations}.

Specifically, with 4 steps, \model 0.6B achieves 5.34 samples/s throughput and 0.32s latency, with an FID of 6.48 and GenEval of 0.76. \model 1.6B has slightly lower throughput (5.20 samples/s) but improves GenEval to 0.77, outperforming larger models like FLUX-schnell (12B), which achieves only 0.5 samples/s with 2.10s latency.
At 2 steps, \model models remain efficient: \model 0.6B reaches 6.46 samples/s with 0.25s latency (FID: 6.54), while \model 1.6B achieves 5.68 samples/s with 0.24s latency (FID: 6.76).
In single-step mode, \model 0.6B achieves 7.22 samples/s throughput and 0.21s latency, maintaining an FID of 7.04 and GenEval of 0.72, comparable to FLUX-schnell but with significantly higher efficiency.

These results demonstrate the practicality of \model for real-time applications, combining fast inference speeds with strong performance metrics.

\input{Tables/loss_cfg_maxt}

\subsection{Analysis}

In this section, we apply a 2-step sampling starting at $t_{max}=\pi/2$ with an intermediate step $t=1.0$.

\noindent \textbf{Schedule Transfer.} To validate the effectiveness of our proposed schedule transfer in~\cref{Sec:trans}, we conduct ablation studies on a flow matching model SANA~\citep{xie2024sana}, comparing its performance with and without schedule transformation to TrigFlow~\citep{lu2024simplifying}. As shown in~\cref{fig:timestep_qknorm_grad}~(d), removing schedule transfer leads to training divergence due to incorrect signals. In contrast, incorporating our schedule transfer enables the model to achieve decent results within 5,000 iterations, demonstrating its crucial role in efficiently adapting flow matching models to TrigFlow-based consistency models.

\noindent \textbf{Influence of CFG Embedding.} To clarify the influence of Classifier-Free Guidance (CFG) embedding in our model, we maintain the setting of incorporating CFG into the teacher model, as established in previous works~\citep{ho2022classifier, luo2023latent, lu2024simplifying}. Specifically, during the training of the student model, we uniformly sample the CFG scale of the teacher model from the set ${4.0, 4.5, 5.0}$. To integrate CFG embedding~\citep{meng2023distillation} into the student model, we add it as an additional condition to the time embedding, multiplying the CFG scale by 0.1 to align with our denser timestep embeddings. We conduct experiments with and without CFG embedding to evaluate its role. As shown in~\cref{tab:cfg_embedding}, incorporating CFG embedding significantly improves CLIP score by 0.94.

\noindent \textbf{Effects of sCM and LADD.} We evaluate the effectiveness of each component by comparing models trained with only the sCM loss or the LADD loss. As shown in~\cref{tab:loss_combination}, training with LADD alone results in instability and suboptimal performance, achieving a higher FID score of 12.20 and a lower CLIP score of 27.00. In contrast, combining both sCM and LADD losses improves model performance, yielding a lower FID score of 8.11 and a higher CLIP score of 28.02, demonstrating their complementary benefits. Using sCM alone achieves a FID score of 8.93 and a CLIP score of 27.51, indicating that while sCM is effective, adding LADD further enhances performance. The weighting ablations for sCM and LADD loss are shown in~\cref{tab:loss_weight}, with additional timestep distribution ablations provided in~\cref{appendix:more ablations}

\noindent \textbf{Additional Max-Time Weighting.} We validate the proposed max-time weighting strategy in LADD (see~\cref{sec:ladd}) through experiments with both sCM and LADD losses. As shown in~\cref{tab:maxT_ladd}, this weighting significantly improves performance. We test the strategy at 0\%, 50\%, and 70\% max-time~($t=\pi/2$) probabilities, finding that 50\% is the best balance, while higher probabilities provide only marginal gains. However, considering the qualitative results,
we finally choose 50\% as the default max-time weighting.

\input{Figures_tex/compare-step}

%% file: Tables/loss_cfg_maxt.tex
\begin{table*}[t]
    \centering
    \begin{minipage}[t]{0.23\textwidth}
        \centering
        \caption{Comparison of loss combination.}
        \label{tab:loss_combination}
        \vspace{-6pt}
        \scalebox{0.75}{
            \begin{tabular}{cc|cc}
                \toprule
                sCM & LADD & FID~$\downarrow$ & CLIP~$\uparrow$ \\
                \midrule
                \checkmark  &               & 8.93  & 27.51 \\ 
                            & \checkmark    & 12.20 & 27.00 \\ 
                \checkmark  & \checkmark    & 8.11  & 28.02 \\
                \bottomrule
            \end{tabular}
        }
    \end{minipage}
    \hfill
    \begin{minipage}[t]{0.24\textwidth}
        \centering
        \small
        \caption{Comparison of CFG training strategies.}
        \label{tab:cfg_embedding}
        \vspace{-6pt}
        \scalebox{0.93}{
            \begin{tabular}{l|cc}
                \toprule
                 CFG Setting & FID~$\downarrow$  & CLIP~$\uparrow$ \\
                \midrule
                w/o Embed           & 9.23  & 27.15 \\
                \textbf{w/ Embed}   & 8.72  & 28.09 \\
                \bottomrule
            \end{tabular}
        }
    \end{minipage}
    \hfill
    \hspace{3pt}
    \begin{minipage}[t]{0.24\textwidth}
        \centering
        \caption{sCM and LADD loss weighting.}
        \label{tab:loss_weight}
        \vspace{-6pt}
        \scalebox{0.75}{
            \begin{tabular}{l|cc}
                \toprule
                    sCM:LADD & FID~$\downarrow$  & CLIP~$\uparrow$ \\
                    \midrule
                    1.0:1.0  & 8.81  & 27.93 \\
                    \textbf{1.0:0.5}  & 8.43  & 27.85 \\
                    1.0:0.1  & 8.90  & 27.76 \\
                \bottomrule
            \end{tabular}
        }
    \end{minipage}
    \hfill
    \hspace{-5pt}
    \begin{minipage}[t]{0.24\textwidth}
        \centering
        \caption{Comparison of max-time weighting strategy.}
        \label{tab:maxT_ladd}
        \vspace{-6pt}
        \scalebox{0.75}{
            \begin{tabular}{c|cc}
                \toprule
                 Max-Time & FID~$\downarrow$  & CLIP~$\uparrow$ \\
                \midrule
                0\%~~ maxT  & 9.44  & 27.65 \\ 
                \textbf{50\% maxT}   & 8.32  & 27.94 \\ 
                70\% maxT   & 8.11  & 28.02 \\ 
                \bottomrule
            \end{tabular}
        }
    \end{minipage}
    \vspace{-1em}
\end{table*}

%% file: Figures_tex/compare-step.tex
\begin{figure}[h]
    \centering
    \includegraphics[width=0.99\linewidth]{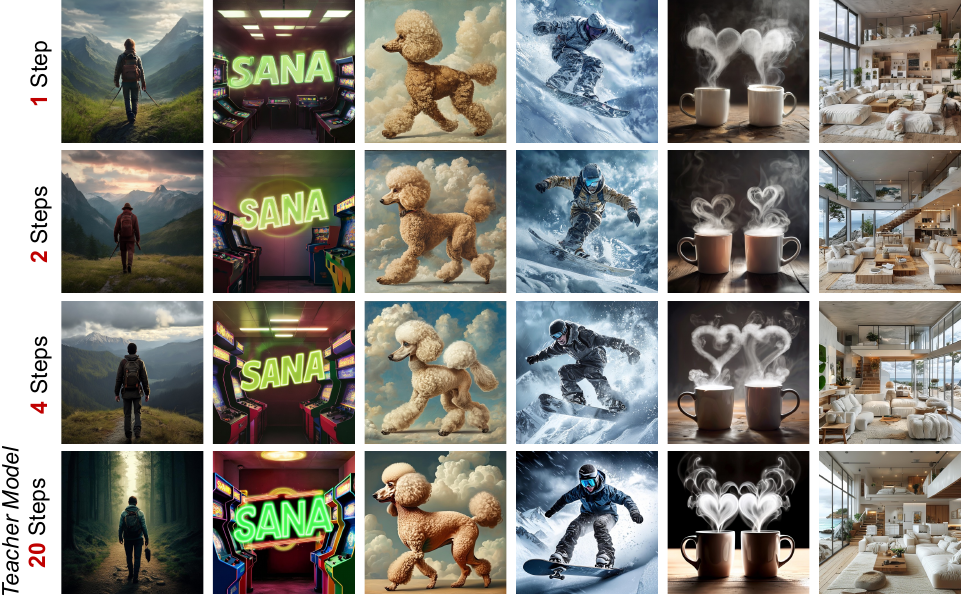}
    \caption{\textbf{Visual comparison among \model with different inference steps and the teacher model SANA.} \model can generate high-quality images with one or two steps and the images can be better when increasing steps. 
    }
    \vspace{-1.5em}
    \label{fig:compare-step}
\end{figure}

%% file: sec/6_related_works.tex
\section{Related Work}

We put a relatively brief overview of related work here, with a more comprehensive version in the appendix. Diffusion models have two primary paradigms for step distillation: trajectory-based and distribution-based methods. Trajectory-based approaches include direct distillation\citep{luhman2021knowledge} and progressive distillation\citep{salimans2022progressive,meng2023distillation}. Consistency models~\citep{song2023consistency} include variants like LCM~\citep{luo2023latent}, CTM~\citep{kim2023consistency}, MCM~\citep{heek2024multistep}, PCM~\citep{wang2024phased}, and sCM~\citep{lu2024simplifying}. Distribution-based methods involve GAN-based distillation~\citep{goodfellow2014generative} and VSD variants~\citep{poole2022dreamfusion,wang2023prolificdreamer,luo2023diff,xie2024distillation,salimans2025multistep}. Recent improvements include adversarial training with DINOv2~\citep{oquab2023dinov2}\citep{sauer2024adversarial}, stabilization of VSD\citep{yin2024one}, and improved algorithms like SID~\citep{zhou2024score} and SIM~\citep{luo2025one}.
In real-time image generation, techniques like PaGoDA\citep{kim2025pagoda} and Imagine-Flash accelerate diffusion inference. Model compression strategies include BitsFusion\citep{sui2024bitsfusion} and Weight Dilation\citep{liu2024dilatequant}. Mobile applications use MobileDiffusion\citep{zhao2024mobilediffusion}, SnapFusion\citep{li2023snapfusion}, and SnapGen\citep{hu2024snapgen}. SVDQuant\citep{li2024svdquant} combined with SANA\cite{xie2024sana} enables fast image generation on consumer GPUs.

%% file: sec/5_conclusion.tex
\section{Conclusion}

In this paper, we introduced SANA-Sprint, an efficient diffusion model for ultra-fast one-step text-to-image generation while preserving multi-step sampling flexibility. By employing a hybrid distillation strategy combining continuous-time consistency distillation~(sCM) and latent adversarial distillation~(LADD), SANA-Sprint achieves SoTA performance with 7.04 FID and 0.72 GenEval in one step, eliminating step-specific training. This unified step-adaptive model enables high-quality 1024$\times$1024 image generation in only 0.1s on H100, setting a new SoTA in speed-quality tradeoffs.

Looking ahead, SANA-Sprint’s instant feedback unlocks real-time interactive applications, transforming diffusion models into responsive creative tools and AIPC. We will open-source our code and models to encourage further exploration in efficient, practical generative AI systems.

%% file: sec/8_acknowledgment.tex
\noindent \textbf{Acknowledgements.}  
We would like to express our heartfelt gratitude to Cheng Lu from OpenAI for his invaluable guidance and insightful discussions on the implementation of the sCM part. We are also deeply thankful to Yujun Lin, Zhekai Zhang, and Muyang Li from MIT for their significant contributions and engaging discussions on the quantization parts, as well as to Lvmin Zhang from Stanford for his expertise and thoughtful input on the ControlNet part. Their collaborative efforts and constructive discussions have been instrumental in shaping this work.

%% file: sec/7_appendix.tex
\renewcommand*{\thesection}{\Alph{section}}

\input{Algorithms/pseudo_code}

\input{Algorithms/transform_algo}

\input{Algorithms/algorithm}

\section{Proof of Proposition 3.1}
\label{app:proof}

Before presenting the formal proof, we first provide some context to understand the necessity of the transformation. In score-based generative models such as diffusion, flow matching, and TrigFlow, denoising is typically performed under certain conditions of data scales and signal-to-noise ratios (SNRs) that match the training setup. However, directly applying flow matching to denoise data generated by TrigFlow is not feasible due to mismatches in time parameterization, SNR, and output necessitating explicit transformations to align the input and output between the two models. The following proof provides the explicit transformation required to connect the TrigFlow-scheduled data to the flow-matching framework.

\input{Algorithms/proof_transformation}

\section{Full Related Work}

\input{sec/6_full_related_works}

\section{More Details}

\subsection{Experimental Setup}\label{app:exp setup}

\paragraph{Model Architecture}
Following the pruning technology in SANA-1.5~\citep{xie2025sana}, our teacher models are fine-tuned from SANA 0.6B and 1.6B, respectively. The architecture, training data, and other hyperparameters remain consistent with SANA-1.5~\citep{xie2025sana}.

\paragraph{Training Details}
We conduct distributed training using PyTorch’s Distributed Data Parallel (DDP) across 32 NVIDIA A100 GPUs on 4 DGX nodes. Our two-phase strategy involves fine-tuning the teacher model with dense time embedding and QK normalization at a learning rate of 2e-5 for 5,000 iterations (global batch size of 1,024), as discussed in \cref{Sec:trans}. 
Then, we perform timestep distillation through the proposed framework at a learning rate of 2e-6 with a global batch size of 512 for 20,000 iterations. As Flash Attention JVP kernel support is not available in PyTorch~\citep{lu2024simplifying}, we retain Linear Attention~\cite{xie2024sana} to auto-compute the JVP.

\paragraph{Evaluation Protocol}
We use multiple metrics: FID, CLIP Score, and GenEval~\citep{ghosh2023geneval}, comparing with state-of-the-art methods. FID and CLIP Score are evaluated on the MJHQ-30K dataset~\citep{li2024playground}. GenEval measures text-image alignment with 553 test prompts, emphasizing its ability to reflect alignment and show improvement potential. We also provide visualizations to compare state-of-the-art methods and highlight our performance.

\subsection{More Ablations}
\label{appendix:more ablations}

\paragraph{Inference Timestep Search}

\cref{fig:inference_steps_search} illustrates the process of timestep optimization for inference across 1, 2, and 4 steps, comparing the performance of 0.6B and 1.6B models in terms of FID~(top row) and CLIP-Scores~(bottom row). The optimization follows a sequential search strategy: first, we determine the optimal $t_{\text{max}}$ for 1-step inference using $\arctan(n/0.5)$, inspired by EDM~\citep{karras2022elucidating}, where $n$ is searched for the maximum timestep. Using this $t_{\text{max}}$, we then search for the intermediate timestep $t_{2nd}$ in 2-step inference. For 4-step inference, the timesteps for the first two steps are fixed to their previously optimized values, while the third~($t_{3rd}$) and fourth timesteps~($t_{4th}$) are searched sequentially. In each case, the x-axis represents the timestep being optimized at the current step, ensuring that earlier steps use their best-found values to maximize overall performance. This hierarchical approach enables efficient timestep selection for multi-step inference settings.

\input{Figures_tex/inference_steps_search}

\input{Figures_tex/scm_mean_std_metrics}

\paragraph{Controlling the sCM Noise Distribution}
In the sCM-only experiments, we investigate the impact of different noise distribution parameter settings on model performance. The noise distribution is defined as $t = \arctan{\left(\frac{e^\tau}{\sigma_d}\right)}$, where $\tau \sim \mathcal{N}(P_{\text{mean}}, P_{\text{std}}^2)$. Starting from the initial parameters $(-0.8, 1.6)$ proposed in sCM~\citep{lu2024simplifying}, we experiment with various mean and standard deviation configurations to evaluate their effects. By tracking FID and CLIP-Score trends over 40k training iterations, we identify $(P_{\text{mean}}, P_{\text{std}}) = (0.0, 1.6)$, represented by the green curve in~\cref{fig:scm_mean_std_metrics}, as the optimal setting. This configuration consistently reduces FID while improving CLIP-Score, resulting in superior generation quality and text-image alignment. 
We also observe that extreme mean values, such as $P_{\text{mean}} = 0.6$ or $P_{\text{mean}} = -0.8$, lead to significant training instability and even failure in some cases. 
Consequently, we adopt $(0.0, 1.6)$ as the default parameter setting.

\paragraph{Controlling the LADD Discriminator Noise Distribution}
Generative features change with the noise level, offering structured feedback at high noise and texture-related feedback at low noise~\citep{sauer2024fast}. We compare the results for different mean and standard deviation settings in $t = \arctan{\left(\frac{e^\tau}{\sigma_d}\right)}$, where $\tau \sim \mathcal{N}(P_{\text{mean}}, P_{\text{std}}^2)$ for LADD discriminator. 
Building on the optimal mean and standard deviation settings $(0.0, 1.6)$ identified for sCM, we further explore the best noise configuration for the LADD's discriminator.
In~\cref{fig:t_mean_std_metrics} and~\cref{tab:t_mean_std_metr_ladd}, we visualize the probability distributions of $t$ sampled under different mean and standard deviation settings, as well as the corresponding FID and CLIP-Score results when applied in the LADD loss. Based on these analyses, we identify $(-0.6, 1.0)$ as the optimal setting, which achieves a more balanced feature distribution across high and low noise levels while maintaining stable training dynamics. Consequently, we adopt $(-0.6, 1.0)$ as the default configuration for LADD loss.

\input{Tables/mean_std_distri}

\subsection{More Qualitative Results}\label{app:qual results}

\paragraph{\model-ControlNet Visualization Images}
In~\cref{fig:controlnet_vis}, we demonstrate the visualization capabilities of our \model-ControlNet, which efficiently achieves impressive results in only \textbf{0.4 seconds} using a \textbf{2-step} generation process, producing high-quality images at a resolution of 1024 $\times$ 1024 pixels. The visualization process begins with an input image, which is processed using a HED detection model to extract the scribe graph. This scribe graph, combined with a given prompt, is used to generate the corresponding image in the second column. The third column presents a blended image that combines the generated image with the scribe graph, highlighting the precise control of the model through boundary alignment. This visualization showcases the model’s ability to accurately interpret prompts and maintain robust control over generated images.

\input{Figures_tex/controlnet_vis}

\input{Figures_tex/controlnet_demo}

\paragraph{More Visualization Images}
In~\cref{fig:generation-results}, we present images generated by our model using various prompts. \model{} showcases comprehensive generation capabilities, including high-fidelity detail rendering, accurate semantic understanding, and reliable text generation, all achieved with only \textbf{2-step sampling}. In particular, the model efficiently produces high-quality images of 1024 $\times$ 1024 pixels in only \textbf{0.24 seconds} on an NVIDIA A100 GPU. The samples demonstrate versatility in various scenarios, from in tricate textures and complex compositions to accurate text rendering, highlighting the robust image quality of the model in both artistic and practical tasks.

\input{Figures_tex/generation-results}

%% file: Algorithms/pseudo_code.tex
\section{Pseudo Code for Training-Free Transformation from Flow to Trigflow}

In this section, we provide a concise implementation of the transformation from a trained flow matching model to a TrigFlow model without requiring additional training. This transformation is based on the theoretical equivalence established in~\cref{app:proof}. The core idea is to first convert the TrigFlow timestep \( t_{\texttt{Trig}} \) to its corresponding flow matching timestep \( t_{\texttt{FM}} \). Then, the input feature \( \bx_{\texttt{Trig}} \) is scaled accordingly to obtain \( \bx_{\texttt{FM}} \). The output of the flow matching model is then transformed using a linear combination to produce the final TrigFlow output. The following pseudo code implements this transformation efficiently, ensuring consistency between the two formulations.



\begin{lstlisting}
class TrigFlowModel(FlowMatchingModel):
    def forward(self, x_trig, t_trig, c):
        t_fm = torch.sin(t_trig) / (torch.cos(t_trig) + torch.sin(t_trig))
        x_fm = x_trig * torch.sqrt(t_fm**2 + (1 - t_fm)**2)

        fm_model_out = super().forward(x_fm, t_fm, c)
        trig_model_out = ((1 - 2 * t_fm) * x_fm + (1 - 2 * t_fm + 2 * t_fm**2) * fm_model_out) / torch.sqrt(t_fm**2 + (1 - t_fm)**2)

        return trig_model_out

\end{lstlisting}

%% file: Algorithms/transform_algo.tex
\section{Transformation Algorithm}
We present an algorithm for training-free transformation from a flow matching model to its TrigFlow counterpart. Given a noisy sample, its corresponding TrigFlow timestep, and a pre-trained flow matching model, the algorithm computes the equivalent flow matching timestep, rescales the input, and applies a deterministic transformation to obtain the TrigFlow output. The detailed procedure is outlined in \cref{alg:1}.
\begin{algorithm}

\caption{Training-Free Transformation to TrigFlow}
\label{alg:1}
\begin{algorithmic}[1]
\State \textbf{Input:} Noisy data $\frac{\bx_{t, \texttt{Trig}}}{\sigma_d}$, timestep $t_\texttt{Trig}$, condition $\boldsymbol{y}$, flow matching model $\boldsymbol{v_{\theta}}(\bx_{t, \texttt{FM}}, t_{\texttt{FM}}, y)$
\State Compute $t_{\texttt{FM}}$ from $t_\texttt{Trig}$ via $t_{\texttt{FM}} = \frac{\sin{(t_{\texttt{Trig}})}}{\sin{(t_{\texttt{Trig}})} + \cos{(t_{\texttt{Trig}})}}$
\State Compute $\bx_{t, \texttt{FM}}$ from $\bx_{t, \texttt{Trig}}$ via $x_{t,\texttt{FM}} = \frac{\bx_{t,\texttt{Trig}}}{\sigma_d}\cdot\sqrt{t_{\texttt{FM}}^2 + (1-t_{\texttt{FM}})^2}$
\State Evaluate $\boldsymbol{v_{\theta}}(\bx_{t, \texttt{FM}}, t_{\texttt{FM}}, y)$
\State Transform the model output via $\small\widehat{\boldsymbol{F_{\theta}}}\left(\frac{\bx_{t,\texttt{Trig}}}{\sigma_d}, t_{\texttt{Trig}}, \boldsymbol{y}\right) = \frac{1}{\sqrt{t_{\texttt{FM}}^2 + (1-t_{\texttt{FM}})^2}} \Big[(1-2t_{\texttt{FM}}) \bx_{t, \texttt{FM}} + (1-2t_{\texttt{FM}} + 2t_{\texttt{FM}}^2) \boldsymbol{v_{\theta}}(\bx_{t, \texttt{FM}}, t_{\texttt{FM}}, \boldsymbol{y})\Big]$
\State \textbf{Output:} Transformed result
\end{algorithmic}
\end{algorithm}

%% file: Algorithms/algorithm.tex
\section{Training Algorithm of \model}
\label{appendix:algorithm}
In this section, we present the detailed training algorithm for \model. To emphasize the differences from the standard sCM training algorithm, we highlight the modified steps in light blue. The following algorithm outlines the complete training procedure, including the key transformations and parameter updates specific to \model.

\newcommand{\algohighlight}[1]{\textcolor{blue}{#1}}
\begin{algorithm}[h]
\caption{Training Algorithm of \model}
\label{alg:sana sprint}
\centering
\begin{algorithmic}[1]
\State \textbf{Input:} dataset $\mathcal{D}$ with std. $\sigma_d = 0.5$, pretrained flow model $\boldsymbol{F}_{\text{pretrain}}$ with parameter $\boldsymbol{\theta}_{\text{pretrain}}$, student model $\boldsymbol{F}_\theta$, \algohighlight{discriminator head $D_{\boldsymbol{\psi}}$}, weighting $w_{\boldsymbol{\phi}}$, learning rate $\eta$, generator distribution ($P_{\text{mean, G}}, P_{\text{std, G}}$), \algohighlight{discriminator distribution ($P_{\text{mean, D}}, P_{\text{std, D}}$)}, constant $c$, warmup iteration $H$, \algohighlight{max-time weighting $p$, condition $\boldsymbol{y}$}.
\State \textbf{Init:} \algohighlight{transform $\boldsymbol{F}_{\text{pretrain}}$ and $\boldsymbol{F}_\theta$ to TrigFlow model using \textbf{Algorithm 1}}, init student model and discriminator backbone with $\boldsymbol{\theta}_{\text{pretrain}}$, $\text{Iters}\gets 0$.
\Repeat
    \State \algohighlight{update discriminator $\boldsymbol{\psi}$}:
    
    \State\hspace{1em} \algohighlight{$\bx_0\sim \mathcal{D}$, $\boldsymbol{z}\sim \mathcal{N}(\mathbf{0},\sigma_d^2\boldsymbol{I})$, $\tau \sim \mathcal{N}(P_{\text{mean, G}}, P^2_{\text{std, G}})$, $t \gets \arctan(\frac{e^\tau}{\sigma_d})$}
    \State\hspace{1em} \algohighlight{if $p > 0$, $\xi\sim \mathrm{U}[0,1]$, $t \gets \frac{\pi}{2}$ if $ \xi < p$}
    \State\hspace{1em} \algohighlight{$\bx_t \gets \cos(t)\bx_0 + \sin(t)\boldsymbol{z}$, $\hat{\bx}_{0}^{\boldsymbol{f_{\theta^-}}} \gets \boldsymbol{f_{\theta^-}}(\bx_t, t, \boldsymbol{y})$}
    \State\hspace{1em} \algohighlight{$\tau \sim \mathcal{N}(P_{\text{mean, D}}, P^2_{\text{std, D}})$, $s \gets \arctan(\frac{e^\tau}{\sigma_d})$}
    \State\hspace{1em} \algohighlight{$\bx_s \gets \cos(s)\bx_0 + \sin(s)\boldsymbol{z}$, $\hat{\bx}_{s}^{\boldsymbol{f_{\theta^-}}} \gets \cos(s)\hat{\bx}_{0}^{\boldsymbol{f_{\theta^-}}}+ \sin(s)\boldsymbol{z}$}
    \State\hspace{1em} \algohighlight{$\small\mathcal{L}_{\text{adv}}^D(\boldsymbol{\psi}) \gets \mathbb{E}_{\bx_0, s}\Big[  \sum_{k} \text{ReLU}\Big(1- D_{\boldsymbol{\psi}, k}(\boldsymbol{F}_{\boldsymbol{\theta}_{\texttt{pre}}, k}(\bx_s, s, \boldsymbol{y}))\Big) \Big] + \mathbb{E}_{\bx_0, s, t}\Big[  \sum_{k} \text{ReLU}\Big(1+ D_{\boldsymbol{\psi}, k}(\boldsymbol{F}_{\boldsymbol{\theta}_{\texttt{pre}}, k}(\hat{\bx}_{s}^{\boldsymbol{f_{\theta^-}}}, s, \boldsymbol{y}))\Big) \Big]$} 
    \State\hspace{1em} \algohighlight{$\boldsymbol{\psi} \gets \boldsymbol{\psi} - \eta \nabla_{\boldsymbol{\psi}}\mathcal{L}_{\text{adv}}^D(\boldsymbol{\psi})$ \Comment{Discriminator step}}
    
    \State\hspace{1em} $\text{Iters}\gets \text{Iters} + 1$

    \State update student model $\boldsymbol{\theta}$ and weighting $\boldsymbol{\phi}$:

    \State\hspace{1em} $\bx_0\sim \mathcal{D}$, $\boldsymbol{z}\sim \mathcal{N}(\mathbf{0},\sigma_d^2\boldsymbol{I})$, $\tau \sim \mathcal{N}(P_{\text{mean, G}}, P^2_{\text{std, G}})$, $t \gets \arctan(\frac{e^\tau}{\sigma_d})$
    \State\hspace{1em} $\bx_t \gets \cos(t)\bx_0 + \sin(t)\boldsymbol{z}$
    
    \State\hspace{1em} $\frac{\mathrm{d} \bx_t}{\mathrm{d} t} \gets \sigma_d\boldsymbol{F}_{\text{pretrain, cfg}}(\frac{\bx_t}{\sigma_d},t)$
    \State\hspace{1em} $r \gets \min(1, \text{Iters} / H)$ \Comment{Tangent warmup}
    \State\hspace{1em} $\boldsymbol{g} \gets -\cos^2(t)(\sigma_d\boldsymbol{F_{\theta^-}} - \frac{\mathrm{d}\bx_t}{\mathrm{d} t}) - r\cdot \cos(t)\sin(t) (\bx_t + \sigma_d \frac{\mathrm{d} \boldsymbol{F_{\theta^-}}}{\mathrm{d} t})$  \Comment{JVP rearrangement}
    \State\hspace{1em} $\boldsymbol{g} \gets \boldsymbol{g} / (\|\boldsymbol{g}\| + c)$ \Comment{Tangent normalization}
    \State\hspace{1em} $\mathcal{L}(\boldsymbol{\theta, \phi}) \gets \frac{e^{w_{\boldsymbol{\phi}}(t)}}{D}\|
        \boldsymbol{F_\theta}(\frac{\bx_t}{\sigma_d},t) - \boldsymbol{F_{\theta^-}}(\frac{\bx_t}{\sigma_d},t)
        - \boldsymbol{g}
    \|_2^2 - w_{\boldsymbol{\phi}}(t)$ \Comment{sCM loss}
    \State\hspace{1em} \algohighlight{if $p > 0$, $\xi\sim \mathrm{U}[0,1]$, $t \gets \frac{\pi}{2}$ if $ \xi < p$}

    \State\hspace{1em} \algohighlight{$\bx_t \gets \cos(t)\bx_0 + \sin(t)\boldsymbol{z}$, $\hat{\bx}_{0}^{\boldsymbol{f_{\theta}}} \gets \boldsymbol{f_{\theta}}(\bx_t, t, \boldsymbol{y})$}

    \State\hspace{1em} \algohighlight{$\hat{\bx}_{s}^{\boldsymbol{f_{\theta}}} \gets \cos(s)\hat{\bx}_{0}^{\boldsymbol{f_{\theta}}}+ \sin(s)\boldsymbol{z}$}
    
    \State\hspace{1em} \algohighlight{$\mathcal{L}(\boldsymbol{\theta, \phi}) \gets \mathcal{L}(\boldsymbol{\theta, \phi}) -\mathbb{E}_{\bx_0, s, t}\Big[  \sum_{k}  D_{\boldsymbol{\psi}, k}(\boldsymbol{F}_{\boldsymbol{\theta}_{\texttt{pre}}, k}(\hat{\bx}_{s}^{\boldsymbol{f_{\theta}}}, s, \boldsymbol{y})) \Big]$} \Comment{GAN loss}
    
    \State\hspace{1em} \algohighlight{$(\boldsymbol{\theta, \phi}) \gets (\boldsymbol{\theta, \phi}) - \eta \nabla_{\theta,\phi}\mathcal{L}(\boldsymbol{\theta, \phi})$} \Comment{Generator step}
    \State\hspace{1em} $\text{Iters}\gets \text{Iters} + 1$
\Until convergence
\end{algorithmic}
\end{algorithm}

%% file: Algorithms/proof_transformation.tex
\begin{proof}

Under the TrigFlow framework, the noisy input sample is given by  
\begin{equation}
\frac{\bx_{t,\texttt{Trig}}}{\sigma_d} = \cos(t_{\texttt{Trig}})\frac{\bx_0}{\sigma_d} + \sin(t_{\texttt{Trig}})\frac{\boldsymbol{z}}{\sigma_d}.
\end{equation}
Since both $\bx_0$ and $\boldsymbol{z}$ originally have a standard deviation of $\sigma_d$, we absorb $\sigma_d$ into these variables so that they are normalized to have a standard deviation of 1. This normalization aligns with the conventions used in flow matching models. Note that the signal-to-noise ratios (SNRs) for flow matching models and TrigFlow models are given by
\begin{equation}
\text{SNR}(t_{\texttt{FM}}) = (\frac{1-t_{\texttt{FM}}}{t_{\texttt{FM}}})^2, \quad \text{SNR}(t_{\texttt{Trig}}) = (\frac{\cos(t_{\texttt{Trig}})}{\sin(t_{\texttt{Trig}})})^2 = (\frac{1}{\tan(t_{\texttt{Trig}})})^2.
\end{equation}
To ensure an equivalent SNR under the flow matching framework, we seek the corresponding time $t_{\texttt{FM}}$ that satisfies:
\begin{equation}
(\frac{1-t_{\texttt{FM}}}{t_{\texttt{FM}}})^2 = (\frac{1}{\tan(t_{\texttt{Trig}})})^2.
\end{equation}
Solving this equation, we obtain the relationship between $t_{\texttt{FM}}$ and $t_{\texttt{Trig}}$:
\begin{equation}
t_{\texttt{FM}} = \frac{\sin{(t_{\texttt{Trig}})}}{\sin{(t_{\texttt{Trig}})} + \cos{(t_{\texttt{Trig}})}}, \quad t_{Trig} = \arctan{(\frac{t_{FM}}{1-t_{FM}})}.
\end{equation}
Under this transformation, the SNRs of $\bx_{t, \texttt{FM}}$ and $\bx_{t, \texttt{Trig}}$ remain equal; however, their scales differ due to the following formulations:
\begin{equation}
\bx_{t, \texttt{FM}} = (1-t_{\texttt{FM}})\bx_0 + t_{\texttt{FM}} \boldsymbol{z}, \quad \bx_{t, \texttt{Trig}} = \cos(t_{\texttt{Trig}})\bx_0 + \sin(t_{\texttt{Trig}}) \boldsymbol{z},
\end{equation}
Since $(1-t_{\texttt{FM}})^2 + t_{\texttt{FM}}^2$ is generally not equal to $\cos^2(t_{\texttt{Trig}}) + \sin^2(t_{\texttt{Trig}}) = 1$ (except when $t_{\texttt{FM}} = 0$ or $1$), a scale adjustment is needed. To align their scales, we introduce a scale factor function $\lambda(t_{\texttt{FM}})$ that satisfies
\begin{equation}
\lambda(t_{\texttt{FM}}) \cdot \cos(t_{\texttt{Trig}}) = (1-t_{\texttt{FM}}), \text{and}\ \lambda(t_{\texttt{FM}}) \cdot \sin(t_{\texttt{Trig}}) = t_{\texttt{FM}},
\end{equation}
Therefore, the scale factor is determined as follows
\begin{equation}
\lambda(t_{\texttt{FM}}) = \frac{1-t_{\texttt{FM}}}{\cos(\arctan{(\frac{t_{\texttt{FM}}}{1-t_{\texttt{FM}}})})} = \frac{t_{\texttt{FM}}}{\sin(\arctan{(\frac{t_{\texttt{FM}}}{1-t_{\texttt{FM}}})})} = \sqrt{t_{\texttt{FM}}^2 + (1-t_{\texttt{FM}})^2}.
\end{equation}
The transformed $\bx_{t,\texttt{Trig}}$ follows the same distribution as the flow matching model's training distribution, achieving our desired objective. Next, we aim to determine the optimal estimator for the TrigFlow model $\boldsymbol{F_{\theta}}$, given $\boldsymbol{v_{\theta}}(\bx_{t, \texttt{FM}}, t_{\texttt{FM}}, \boldsymbol{y})$. We first consider an ideal scenario where the model's capacity is sufficiently large. In this case, the flow matching model reaches its optimal solution:
\begin{equation}
\boldsymbol{v}^*(\bx_{t, \texttt{FM}}, t_{\texttt{FM}}, \boldsymbol{y}) = \mathbb{E}[\boldsymbol{z}-\bx_0 | \bx_{t_{\texttt{FM}}}, \boldsymbol{y}],
\end{equation}
as the conditional expectation minimizes the mean squared error (MSE) loss. Similarly, the optimal solution of the TrigFlow model is given by
\begin{equation}
\boldsymbol{F}^*(\bx_{t, \texttt{Trig}}, t_{\texttt{Trig}}, \boldsymbol{y}) = \mathbb{E}[\cos{(t_{\texttt{Trig}})}\boldsymbol{z}-\sin{(t_{\texttt{Trig}})}\bx_0 | \bx_{t_{\texttt{Trig}}}, \boldsymbol{y}].
\end{equation}
Noting that
\begin{equation}
\cos{(t_{\texttt{Trig}})} = \frac{1-t_{\texttt{FM}}}{ \sqrt{t_{\texttt{FM}}^2 + (1-t_{\texttt{FM}})^2}}, \quad \sin{(t_{\texttt{Trig}})} = \frac{t_{\texttt{FM}}}{ \sqrt{t_{\texttt{FM}}^2 + (1-t_{\texttt{FM}})^2}},
\end{equation}
we leverage the linearity of conditional expectation to derive
\begin{equation}
\begin{aligned}
&\frac{1-2t_{\texttt{FM}}}{\sqrt{t_{\texttt{FM}}^2 + (1-t_{\texttt{FM}})^2}}\cdot \bx_{t_{\texttt{FM}}}
+ \frac{1-2t_{\texttt{FM}}+2t_{\texttt{FM}}^2}{\sqrt{t_{\texttt{FM}}^2 + (1-t_{\texttt{FM}})^2}}\mathbb{E}[\boldsymbol{z}-\bx_0 | \bx_{t_{\texttt{FM}}}, \boldsymbol{y}]\\
=& \frac{1-2t_{\texttt{FM}}}{\sqrt{t_{\texttt{FM}}^2 + (1-t_{\texttt{FM}})^2}} \mathbb{E}[(1-t_{\texttt{FM}})\cdot\bx_0 + t_{\texttt{FM}}\cdot\boldsymbol{z}| \bx_{t_{\texttt{FM}}}, \boldsymbol{y}]
+ \frac{1-2t_{\texttt{FM}}+2t_{\texttt{FM}}^2}{\sqrt{t_{\texttt{FM}}^2 + (1-t_{\texttt{FM}})^2}}\mathbb{E}[\boldsymbol{z}-\bx_0 | \bx_{t_{\texttt{FM}}}, \boldsymbol{y}]\\
=& \mathbb{E}[ \frac{1-t_{\texttt{FM}}}{\sqrt{t_{\texttt{FM}}^2 + (1-t_{\texttt{FM}})^2}} \boldsymbol{z} - \frac{t_{\texttt{FM}}}{\sqrt{t_{\texttt{FM}}^2 + (1-t_{\texttt{FM}})^2}}\bx_0| \bx_{t_{\texttt{FM}}}, \boldsymbol{y}]\\
\\
=& \mathbb{E}[ \cos{(t_{\texttt{Trig}})}\boldsymbol{z} - \sin{(t_{\texttt{Trig}})}\bx_0 | x_{t_{\texttt{Trig}}}, \boldsymbol{y}].
\end{aligned}
\end{equation}
Consequently, we obtain
\begin{equation}
\boldsymbol{F}^*(\bx_{t, \texttt{Trig}}, t_{\texttt{Trig}}, \boldsymbol{y})
= \frac{1}{\sqrt{t_{\texttt{FM}}^2 + (1-t_{\texttt{FM}})^2}} \Big[
  (1-2t_{\texttt{FM}}) \bx_{t, \texttt{FM}} + (1-2t_{\texttt{FM}} + 2t_{\texttt{FM}}^2) \boldsymbol{v}^*(\bx_{t, \texttt{FM}}, t_{\texttt{FM}}, \boldsymbol{y})
\Big].
\end{equation}
Next, we consider a more realistic scenario where the model's capacity is limited, leading to the learned velocity field
\begin{equation}
\boldsymbol{v_{\theta^*}}(\bx_{t, \texttt{FM}}, t_{\texttt{FM}}, \boldsymbol{y}) = \min_{\boldsymbol{\theta}} \mathbb{E}_{\bx_0, \bz, t}[w(t)\| \boldsymbol{v_\theta}(\bx_{t, \texttt{FM}}, t_{\texttt{FM}}, \boldsymbol{y}) - (\bz-\bx_0) \|^2].
\end{equation}
Under our parameterization, training the TrigFlow model amounts to minimizing
\begin{equation}\footnotesize
\begin{aligned}
&\min_{\boldsymbol{\theta}} \mathbb{E}_{\bx_0, \bz, t}\left[ \left\| 
\frac{1-2t_{\texttt{FM}}}{\sqrt{t_{\texttt{FM}}^2 + (1-t_{\texttt{FM}})^2}}\cdot \bx_{t_{\texttt{FM}}}
+ \frac{1-2t_{\texttt{FM}}+2t_{\texttt{FM}}^2}{\sqrt{t_{\texttt{FM}}^2 + (1-t_{\texttt{FM}})^2}}\boldsymbol{v_\theta}(\bx_{t, \texttt{FM}}, t_{\texttt{FM}}, \boldsymbol{y}) - \left(\cos(t_{\texttt{Trig}}) \boldsymbol{z} - \sin(t_{\texttt{Trig}})\bx_0 \right)
 \right\|^2\right]\\
=&\min_{\boldsymbol{\theta}} \mathbb{E}_{\bx_0, \bz, t}\left[ \left\| 
\frac{1-2t_{\texttt{FM}}}{\sqrt{t_{\texttt{FM}}^2 + (1-t_{\texttt{FM}})^2}}\cdot \bx_{t_{\texttt{FM}}}
+ \frac{1-2t_{\texttt{FM}}+2t_{\texttt{FM}}^2}{\sqrt{t_{\texttt{FM}}^2 + (1-t_{\texttt{FM}})^2}}\boldsymbol{v_\theta}(\bx_{t, \texttt{FM}}, t_{\texttt{FM}}, \boldsymbol{y}) - \left(\frac{1-t_{\texttt{FM}}}{ \sqrt{t_{\texttt{FM}}^2 + (1-t_{\texttt{FM}})^2}} \boldsymbol{z} - \frac{t_{\texttt{FM}}}{ \sqrt{t_{\texttt{FM}}^2 + (1-t_{\texttt{FM}})^2}}\bx_0 \right)
 \right\|^2\right]\\
\end{aligned}
\end{equation}
Substituting $\bx_{t, \texttt{FM}} = (1-t_{\texttt{FM}})\bx_0 + t_{\texttt{FM}} \boldsymbol{z}$, the above expression simplifies to
\begin{equation}\footnotesize
\begin{aligned}
&\min_{\boldsymbol{\theta}} \mathbb{E}_{\bx_0, \bz, t}\left[ \left\| 
\frac{1-2t_{\texttt{FM}}}{\sqrt{t_{\texttt{FM}}^2 + (1-t_{\texttt{FM}})^2}}\cdot \bx_{t_{\texttt{FM}}}
+ \frac{1-2t_{\texttt{FM}}+2t_{\texttt{FM}}^2}{\sqrt{t_{\texttt{FM}}^2 + (1-t_{\texttt{FM}})^2}}\boldsymbol{v_\theta}(\bx_{t, \texttt{FM}}, t_{\texttt{FM}}, \boldsymbol{y}) - \left(\frac{1-t_{\texttt{FM}}}{ \sqrt{t_{\texttt{FM}}^2 + (1-t_{\texttt{FM}})^2}} \boldsymbol{z} - \frac{t_{\texttt{FM}}}{ \sqrt{t_{\texttt{FM}}^2 + (1-t_{\texttt{FM}})^2}}\bx_0 \right)
 \right\|^2\right]\\
=& \min_{\boldsymbol{\theta}} \mathbb{E}_{\bx_0, \bz, t}\left[ \left\| 
\frac{t_{\texttt{FM}}^2 + (1-t_{\texttt{FM}})^2}{\sqrt{t_{\texttt{FM}}^2 + (1-t_{\texttt{FM}})^2}}\boldsymbol{v_\theta}(\bx_{t, \texttt{FM}}, t_{\texttt{FM}}, \boldsymbol{y}) - \left(\frac{t_{\texttt{FM}}^2 + (1-t_{\texttt{FM}})^2}{ \sqrt{t_{\texttt{FM}}^2 + (1-t_{\texttt{FM}})^2}} \boldsymbol{z} - \frac{t_{\texttt{FM}}^2 + (1-t_{\texttt{FM}})^2}{ \sqrt{t_{\texttt{FM}}^2 + (1-t_{\texttt{FM}})^2}}\bx_0 \right)
 \right\|^2\right]\\
=& \min_{\boldsymbol{\theta}} \mathbb{E}_{\bx_0, \bz, t}\left[\left(t_{\texttt{FM}}^2 + (1-t_{\texttt{FM}})^2\right) \left\| 
\boldsymbol{v_\theta}(\bx_{t, \texttt{FM}}, t_{\texttt{FM}}, \boldsymbol{y}) - \left( \boldsymbol{z} - \bx_0 \right)
 \right\|^2\right]\\
\end{aligned}
\end{equation}
Thus, training the TrigFlow model with our parameterization is equivalent to training the flow matching model, apart from differences in the loss weighting function $w(t)$ and the timestep sampling distribution $p(t)$.

\end{proof}

%% file: sec/6_full_related_works.tex
\paragraph{Text to Image Generation}
Text-to-image generation has experienced transformative advancements in both efficiency and model design. The field gained early traction with Stable Diffusion~\citep{rombach2022high}, which set the stage for scalable high-resolution synthesis. A pivotal shift occurred with Diffusion Transformers (DiT)\citep{Peebles2022DiT}, which replaced conventional U-Net architectures with transformer-based designs, unlocking improved scalability and computational efficiency. Building on this innovation, PixArt-$\alpha$\citep{chenpixart} demonstrated competitive image quality while slashing training costs to only 10.8\% of those required by Stable Diffusion v1.5~\citep{rombach2022high}.
Recent breakthroughs have further pushed the boundaries of compositional generation. Large-scale models like FLUX~\citep{FLUX} and Stable Diffusion 3~\citep{sd3} have scaled up to ultra-high-resolution synthesis and introduced multi-modal capabilities through frameworks such as the Multi-modal Diffusion Transformer (MM-DiT)\citep{esser2024scaling}. Playground v3\citep{liu2024playground} achieved state-of-the-art image quality by seamlessly integrating diffusion models with Large Language Models (LLMs)\citep{dubey2024llama}, while PixArt-$\Sigma$\citep{chenpixartsigma} showcased direct 4K image generation using a compact 0.6B parameter model, emphasizing computational efficiency alongside high-quality outputs.
Efficiency-driven innovations have also gained momentum. SANA~\citep{xie2024sana} introduced high-resolution synthesis capabilities through deep compression autoencoding~\citep{chen2024deep} and linear attention mechanisms, enabling deployment on consumer-grade hardware like laptop GPUs. Additionally, advancements in linear attention mechanisms for class-conditional generation~\citep{cai2024condition, zhu2024dig}, diffusion models without attention~\citep{yan2024diffusion, teng2024dim}, and cascade structures~\citep{pernias2023wuerstchen, ren2024ultrapixel, tian2024u} have further optimized computational requirements while maintaining performance. These developments collectively underscore the field’s rapid evolution toward more accessible, efficient, and versatile text-to-image generation technologies.

\paragraph{Diffusion Model Step Distillations}
Current methodologies primarily coalesce into two dominant paradigms: (1) trajectory-based distillation. Direct Distillation~\citep{luhman2021knowledge} directly learns noise-image mapping given by PF-ODE. Progressive Distillation~\citep{salimans2022progressive,meng2023distillation} makes the learning progress easier by progressively enlarging subintervals on the ODE trajectory. Consistency Models (CMs)~\citep{song2023consistency} (e.g. LCM~\citep{luo2023latent}, CTM~\citep{kim2023consistency}, MCM~\citep{heek2024multistep}, PCM~\citep{wang2024phased}, sCM~\citep{lu2024simplifying}) predict the solution $\bx_0$ of the PF-ODE given $\bx_t$ via self-consistency. (2) distribution-based distillation. It can be further divided into GAN~\citep{goodfellow2014generative}-based distillation and its variational score distillation (VSD) variants~\citep{poole2022dreamfusion,wang2023prolificdreamer,luo2023diff,xie2024distillation,salimans2025multistep}. ADD~\citep{sauer2024adversarial} explored distilling diffusion models using adversarial training with pretrained feature extractor like DINOv2~\citep{oquab2023dinov2} in pixel space. LADD~\citep{sauer2024fast} further utilize teacher diffusion models as feature extractors enabling direct discrimination in latent space, drastically saving the computation and GPU memories. \citep{yin2024one} stabilize VSD with regression loss. SID~\citep{zhou2024score} and SIM~\citep{luo2025one} propose improved algorithms for VSD.

\paragraph{Real-Time Image Generation}
Recent advancements in real-time image generation have focused on improving the efficiency and quality of diffusion models. PaGoDA~\citep{kim2025pagoda} introduces a progressive approach for one-step generation across resolutions. Imagine-Flash also uses a backward distillation to accelerate diffusion inference. In model compression, BitsFusion~\citep{sui2024bitsfusion} quantizes Stable Diffusion's UNet to 1.99 bits, and Weight Dilation~\citep{liu2024dilatequant} presents DilateQuant for enhanced performance. For mobile applications, MobileDiffusion~\citep{zhao2024mobilediffusion} achieves sub-second generation times, with SnapFusion~\citep{li2023snapfusion} and SnapGen~\citep{hu2024snapgen} enabling 1024x1024 pixel image generation in about 1.4 seconds. SVDQuant~\citep{li2024svdquant} introduces 4-bit quantization for diffusion models, and when combined with SANA~\citep{xie2024sana}, enables fast generation of high-quality images on consumer GPUs, bridging the gap between model performance and real-time applications.

%% file: Figures_tex/inference_steps_search.tex
\begin{figure}[ht]
    \centering
    \begin{minipage}[t]{\textwidth}
        \centering
        \includegraphics[width=0.88\linewidth]{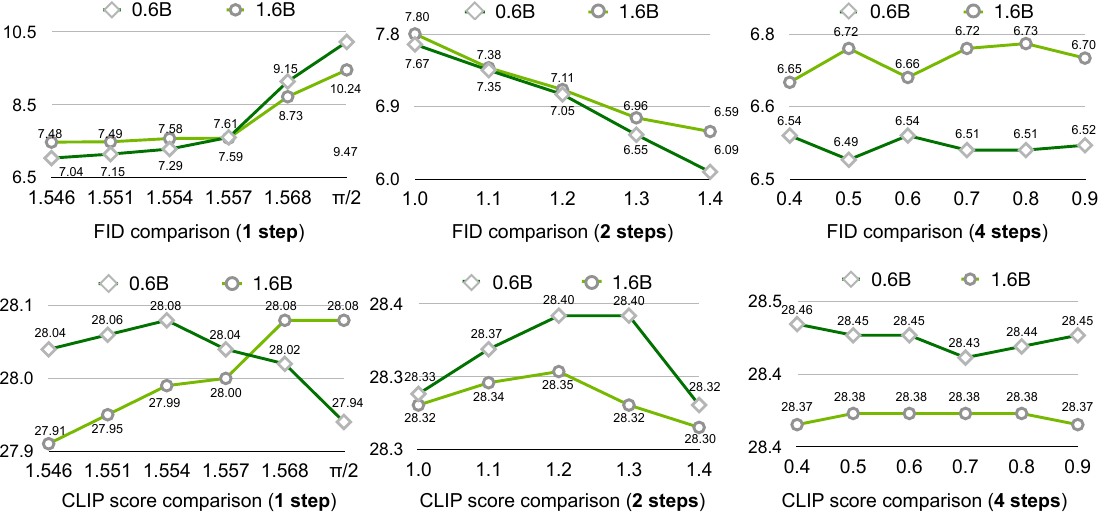}
        \vspace{-0.5em}
        \caption{
        \textbf{Inference timesteps search.} This figure illustrates the performance of timesteps search for achieving optimal results during inference with 0.6B and 1.6B models. The subplots compare FID~(top row) and CLIP-Score~(bottom row) across different timesteps for 1-step, 2-step, and 4-step inference settings. The x-axis represents the timestep being searched at the current step; for multi-step settings (e.g., 4 steps), the timesteps for earlier steps are fixed to their previously optimized values.
        }
        \label{fig:inference_steps_search}
    \end{minipage}

    \vspace{1em}

    \begin{minipage}[t]{\textwidth}
        \centering
        \captionof{table}{Inference timestep settings for both \model 0.6B and 1.6B models.}
        \label{tab:inference_timestep_setting}
        \vspace{-0.5em}
        \scalebox{0.95}{
            \begin{tabular}{c|c|c|c}
                \toprule
                 & \textbf{1 step} & \textbf{2 steps} & \textbf{4 steps} \\
                \midrule
                Timestep T & $[\pi/2, 0.0]$ & $[\arctan(200/0.5), 1.3, 0.0]$ & $[\arctan(200/0.5), 1.3, 1.1, 0.6, 0.0]$ \\
                \bottomrule
            \end{tabular}
        }
    \end{minipage}
\end{figure}

%% file: Figures_tex/scm_mean_std_metrics.tex
\begin{figure*}[h]
    \centering
    \includegraphics[width=0.85\linewidth]{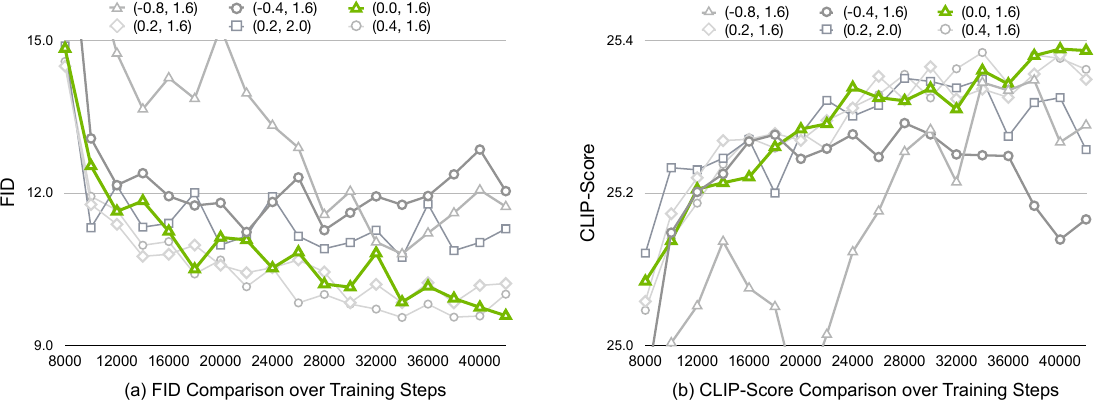}
    \caption{\textbf{Controlling the sCM noise distribution.} This figure compares FID and CLIP-Score across different noise distribution settings over 40k training steps in sCM-only experiments. The green curve $(P_{\text{mean}}, P_{\text{std}}) = (0.0, 1.6)$ demonstrates optimal performance, achieving stable training dynamics and superior generation quality.
    }
    \label{fig:scm_mean_std_metrics}
    \vspace{-1em}
\end{figure*}

%% file: Tables/mean_std_distri.tex
\begin{figure}[h]
    \centering
    \begin{minipage}[t]{0.48\textwidth}
        \centering
        \vspace{0pt}
        \includegraphics[width=0.9\linewidth]{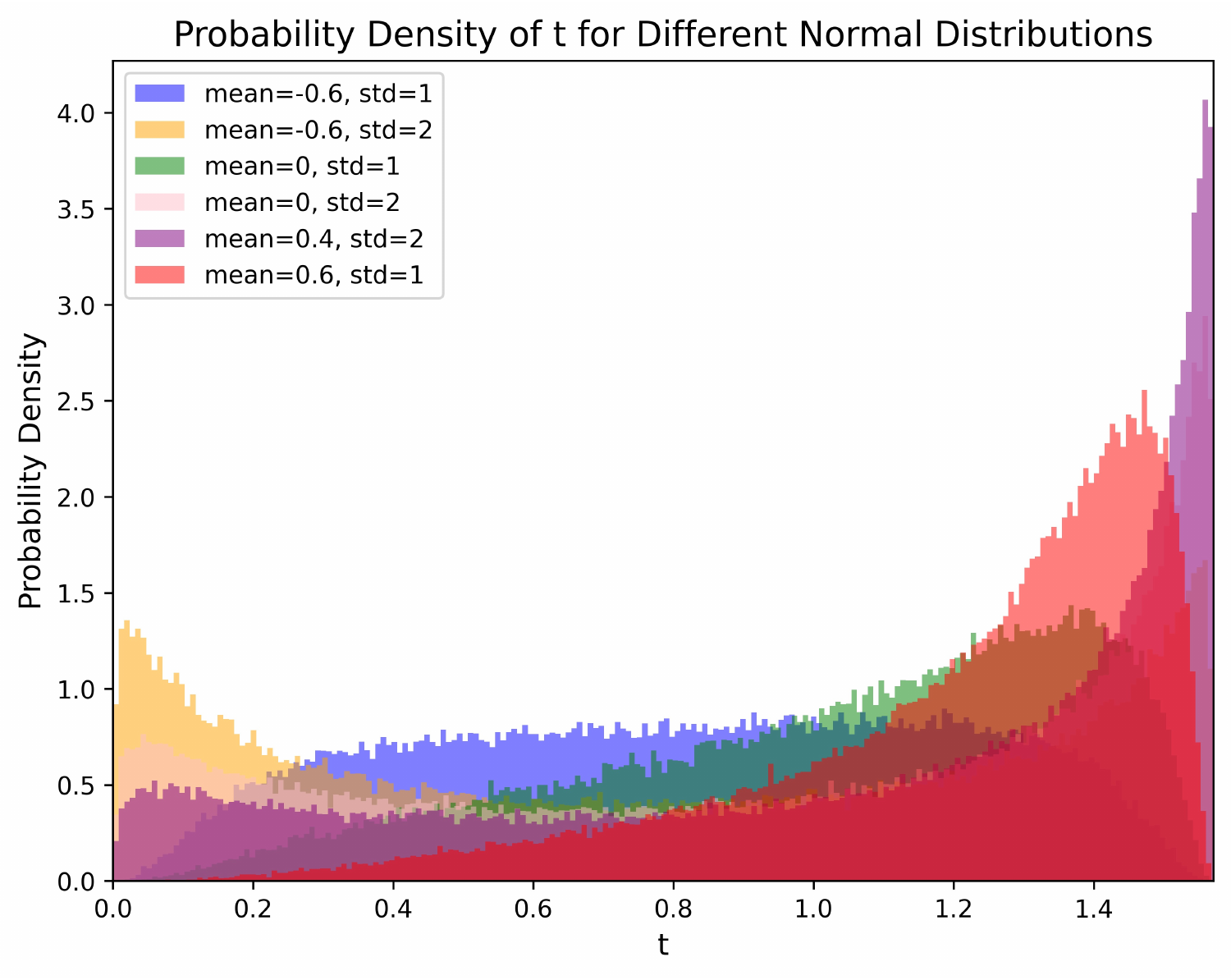}
        \caption{\textbf{Controlling the LADD noise distribution.} We vary the parameters of a logit-normal distribution for biasing the sampling of the LADD teacher noise level. When biasing towards very high noise levels (m = 0.4, s = 2), we observe unstable training.
        }
        \label{fig:t_mean_std_metrics}
    \end{minipage}
    \hfill
    \begin{minipage}[t]{0.5\textwidth}
        \centering
        \vspace{0pt}
        \captionof{table}{Comparison of different noise distributions for LADD loss.}
        \label{tab:t_mean_std_metr_ladd}
        \vspace{-6pt}
        \scalebox{0.95}{
            \begin{tabular}{c|cc}
                \toprule
                Mean, Std & FID~$\downarrow$ & CLIP~$\uparrow$ \\
            \midrule
                \textbf{(-0.6, 1.0)}     & 9.48      & 28.08     \\ 
                (-0.6, 2.0)     & 10.36     & 28.03     \\ 
                (0.0, 1.0)      & 13.11     & 27.18     \\
                (0.0, 2.0)      & 11.25     & 27.96     \\
                (0.4, 2.0)      & 9.77      & 28.00     \\
                (0.6, 1.0)      & 12.85     & 27.32     \\
                \bottomrule
            \end{tabular}
        }
    \end{minipage}
\end{figure}

%% file: Figures_tex/controlnet_vis.tex
\begin{figure*}[th]
    \centering
    \includegraphics[width=0.98\linewidth]{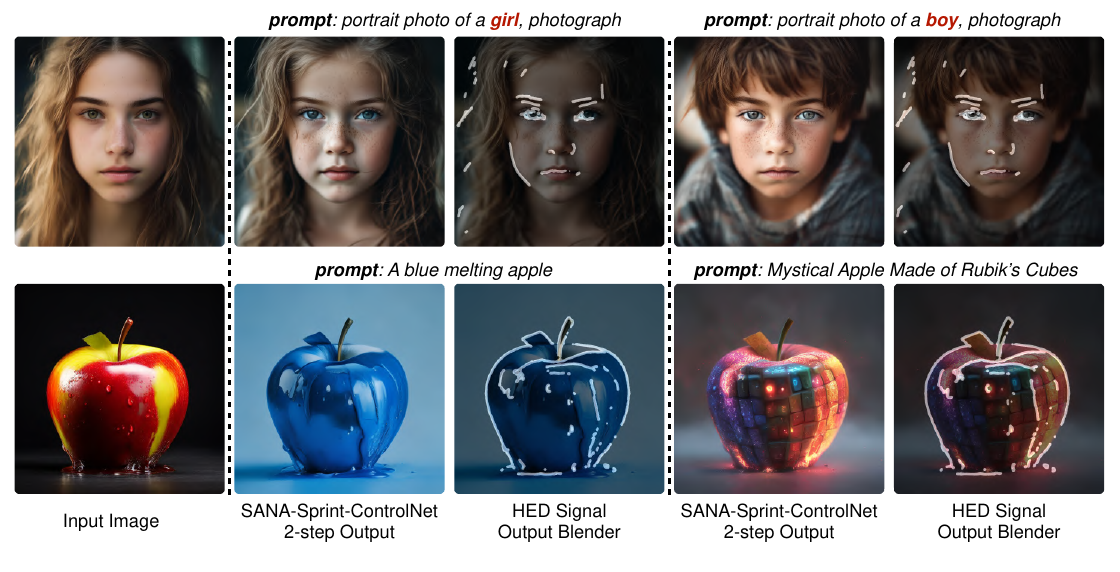}
    \vspace{-1.7em}
    \caption{\textbf{Visualization of \model-ControlNet’s capabilities.} The model outputs high-quality images of 1024 $\times$ 1024 pixels in only \textbf{2 steps} and \textbf{0.3 seconds} on an NVIDIA H100 GPU. The process involves processing the input image (first column) to extract a scribe graph, which, along with a prompt, generates an image (second column). The blended image (third column) highlights precise boundary alignment and control, demonstrating the model’s robust control capabilities. 
    }
    \label{fig:controlnet_vis}
    \vspace{-1em}
\end{figure*}

%% file: Figures_tex/controlnet_demo.tex
\begin{figure*}[t]
    \centering
    \includegraphics[width=0.98\linewidth]{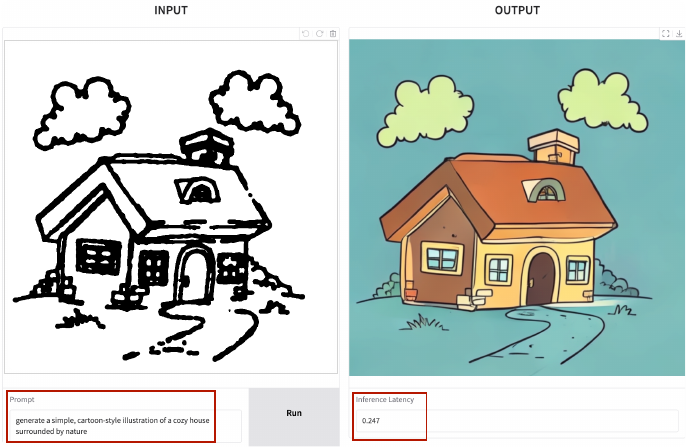}
    \vspace{-0.7em}
    \caption{\textbf{ControlNet Demo: Hand-Crafted Scribble to Stunning Image.} 
    \textbf{Left}: A hand-crafted scribble created with a brush. \textbf{Right}: The result generated by the Sana-Sprint-ControlNet model, strictly following the scribble and prompt. \textbf{Inference Latency}: The model achieves remarkable speed, generating the 1024 $\times$ 1024 images in only \textbf{1 step} and \textbf{0.25 seconds} on H100 GPU, as shown in the right red box. This demo showcases the model’s exceptional control and efficiency, adhering closely to the user’s input while producing visually appealing results.
    }
    \label{fig:controlnetGUI}
\end{figure*}

%% file: Figures_tex/generation-results.tex
\begin{figure*}[t]
    \centering
    \includegraphics[width=0.99\linewidth]{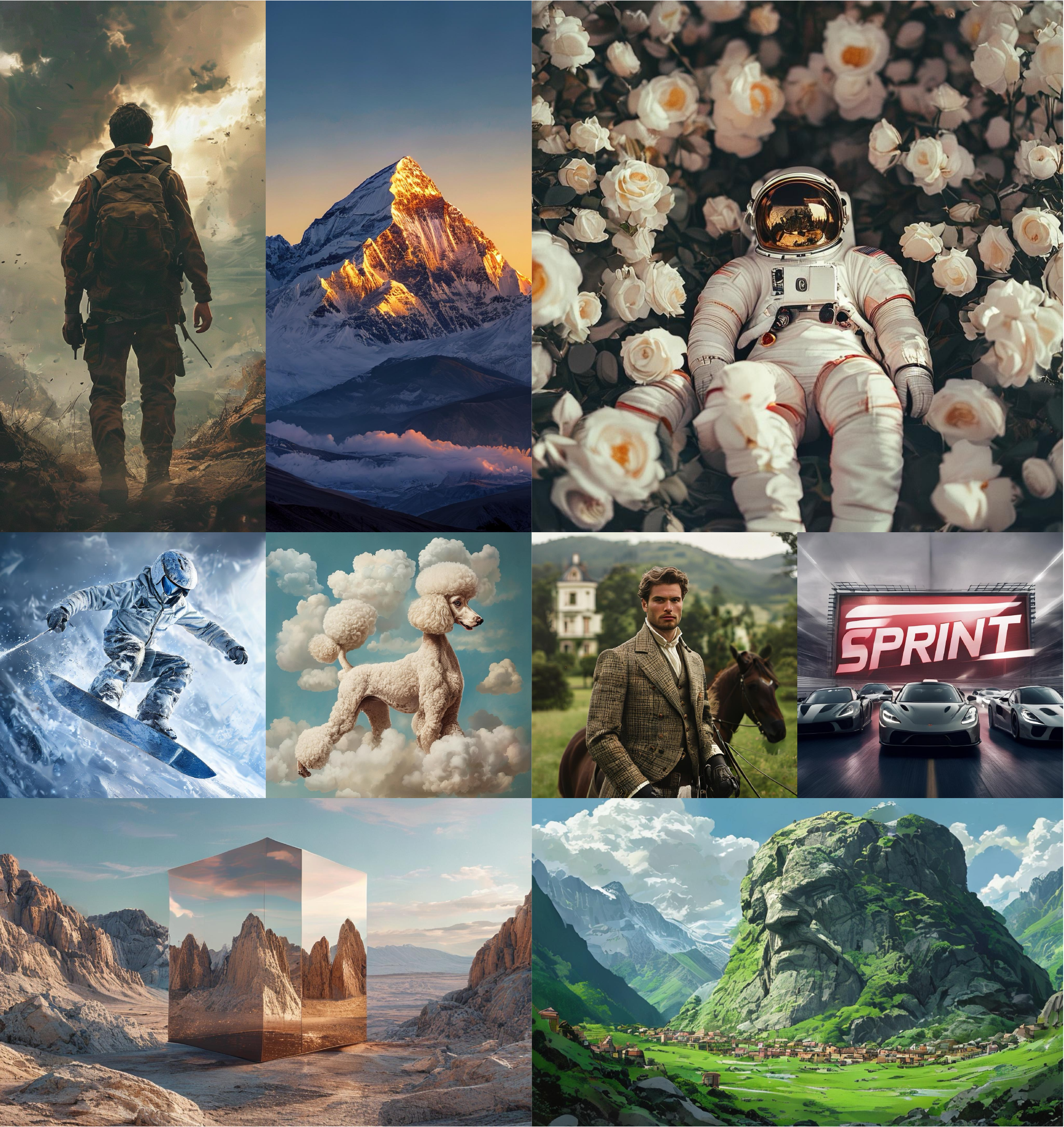}
    \caption{\textbf{Generated images with \model.} The model outputs high-quality images of 1024 $\times$ 1024 pixels in \textbf{2 steps} and \textbf{0.24 seconds} on an NVIDIA A100 GPU, showcasing comprehensive generation capabilities with high-fidelity details and accurate text rendering, handling diverse scenarios with robust image quality.
    }
    \label{fig:generation-results}
\end{figure*}